%% file: paper.tex
\documentclass[submission, PhysCommReports]{SciPost}

\input{incl_settings.tex}

\input{incl_shortcuts.tex}

\usepackage{longtable}
\graphicspath{{./figs/}}

\begin{document}

\begin{center}{
\Large \textbf{\color{scipostdeepblue}{
VERaiPHY -- Validation \& Evaluation for Robust AI in PHYsics}}}\\[0.3cm]
{\large Introduction to the VERaiPHY Initiative}
\end{center}

\begin{center}
Gaia Grosso\textsuperscript{1--3},
Ramon Winterhalder\textsuperscript{4},\\
Lydia Brenner\textsuperscript{5},
Louis Lyons\textsuperscript{6,7},
and Tilman Plehn\textsuperscript{8,9}
\end{center}

\begin{center}
{\bf 1} NSF AI Institute for Artificial Intelligence and Fundamental Interactions, Cambridge, MA\\
{\bf 2} MIT Laboratory for Nuclear Science, Cambridge, USA\\
{\bf 3} School of Engineering and Applied Sciences, Harvard University, Cambridge, USA\\
{\bf 4} TIFLab, Universit\`a degli Studi di Milano \& INFN Sezione di Milano, Italy\\
{\bf 5} Nikhef, Amsterdam, The Netherlands\\
{\bf 6} Department of Physics, Blackett Laboratory, Imperial College London, UK\\
{\bf 7} Department of Physics, Oxford University, UK\\
{\bf 8} Institut f\"ur Theoretische Physik, Universit\"at Heidelberg, Germany\\
{\bf 9} Interdisciplinary Center for Scientific Computing (IWR), Universit\"at Heidelberg, Germany
\end{center}

\begin{center}
\today
\end{center}


\section*{\color{scipostdeepblue}{Abstract}}
{\bf 
Modern machine learning is leading to substantial gains in precision, flexibility, and computational efficiency in fundamental physics. 
Statistical validation, uncertainty quantification, and robustness assessment are less systematically addressed.
The VERaiPHY initiative (Validation \& Evaluation for Robust AI in PHYsics) is a series of articles developed within the PHYSTAT programme, aimed at establishing statistical standards for the development, evaluation, and deployment of ML techniques.
Each article focuses on a specific methodological domain from a statistics perspective and clarifies statistical questions, tests, and the interpretation of results.
This opening article establishes the probabilistic, statistical, and machine learning foundations that the later contributions assume, together with the notation used throughout.
}

\clearpage
\vspace{10pt}
\noindent\rule{\textwidth}{1pt}
\tableofcontents\thispagestyle{fancy}
\noindent\rule{\textwidth}{1pt}
\vspace{10pt}

\clearpage
\section{Introduction}

Over the past decade, modern machine learning (ML) has transformed nearly every aspect of fundamental physics research.
Fundamental physics presents a particularly fertile ground for scientific AI methods: its datasets are vast and high-dimensional, its complexity demands sophisticated analysis, and the availability of first-principle simulations enables rigorous validation.
Applications start with designing and running experiments and detectors. 
They are changing how we think of anomaly detection, triggering, data acquisition, calibration, and object reconstruction. 
In data analysis, ML opens new avenues for object identification, regression tasks, and uncertainty modelling.
For theory predictions, ML methods enhance analytic manipulations, numerical evaluations of quantum field theory and general relativity predictions, and sophisticated multi-step simulation pipelines.
Optimal simulation-based inference and unbinned high-dimensional unfolding would not be possible without ML. 
In all these applications, ML methods often dramatically outperform traditional techniques in both accuracy and computational efficiency.

However, accurately learned central values do not imply useful measurements or predictions without  reliable uncertainty estimates and improved performance in an ML sense does not automatically lead to improved scientific results. A narrow uncertainty band is only useful if it can be shown to be calibrated; a powerful classifier may induce bias and may not generalise from simulated training data to actual data; a generative model may reproduce distributions without preserving statistical guarantees. 
As ML provides central components of simulation and analysis pipelines, the question is no longer whether they work operationally, but whether they are validated statistically.

Crucially, ML methods in fundamental physics should not replace first-principle predictions and reduce proper analyses to data-driven self-consistency tests. 
A network reproducing data perfectly may implicitly encode the Higgs self-coupling, the value of the Hubble constant, or the existence of a dark matter particle.
For it to be useful for fundamental physics, results have to be at a high level of theoretical abstraction: a Lagrangian, Hamiltonian, or action encoding the fundamental constituents, symmetries, and interactions of 
Nature at a given scale. This understanding can be inferred in a statistically rigorous manner.
In ML terms, fundamental physics requires interpretable and statistically 
validated latent representations of physical laws and their parameters, combining insights from particle physics, astrophysics, and cosmology.

These pressing questions motivated the PHYSTAT meeting “Statistics meets Machine Learning” in September 2024\footnote{\url{https://indico.cern.ch/event/1407421/}}. 
The PHYSTAT workshop series began in January 2000 at CERN. 
Since then, PHYSTAT meetings have provided a forum for discussing statistical challenges arising in particle physics, with increasing engagement from astrophysics and cosmology. 
In 2019, PHYSTAT Seminars were introduced to extend this dialogue, followed more recently by PHYSTAT Informal Reviews, which bring statisticians and physicists together to examine specific statistical procedures, often originally developed within particle physics, to assess their broader validity and limitations. 
The first PHYSTAT Statistics School further expanded this educational and interdisciplinary effort. 
Details of these activities can be found on the PHYSTAT homepage\footnote{\url{https://phystat.github.io/Website/}}.

The 2024 ML-PHYSTAT meeting highlighted many promising and even transformative developments, but it also brought into focus a recurring question: \emph{How can we relate the powerful new ML methods to statistical tests?} We urgently need to address this question in fundamental physics if we want to (i) benefit from the transformative power of ML methods, (ii) extract all information from our vast amounts of complex data, and (iii) provide relevant training to our young scientists. The VERaiPHY article series addresses this gap. Individual articles on dedicated ML topics examine a specific methodological domain from a statistics-first perspective:
\begin{enumerate}
    \item Statistical Properties of Training \& Generalisation~\cite{Lavie:2026psw}
    \item Uncertainty Quantification~\cite{Haussmann:2026gbi}
    \item Model Robustness~\cite{Cruz-Martinez:2026apu}
    \item Inference \& Parameter Estimation~\cite{Dax:2026qsm}
    \item Generative Models \& Statistical Validation~\cite{Diefenbacher:2026kki}
    \item Hypothesis Testing \& Statistical Discovery~\cite{Amram:2026vkc}
    \item Symmetries \& Inductive Biases~\cite{VERaiPHY_sym}
    \item Interpretability \& Explainability~\cite{Gambhir:2026cly}
    \item Representation Learning~\cite{VERaiPHY_rep}
    \item Future Directions \& Open Challenges~\cite{Grosso:2026fgo}
\end{enumerate}
Each article is authored by early-career researchers with complementary expertise spanning particle physics, astronomy, cosmology, statistics, and computer science. 
In some cases they are supported by senior advisors.
Taken together, the articles establish a framework for evaluating ML methods in physics in terms of their statistical framework, validity, and robustness. 
In doing so, VERaiPHY seeks to contribute to a more rigorous and sustainable integration of AI methods into the scientific practice of fundamental physics.
 
This introduction should serve as a standing reference throughout the VERaiPHY series rather than as a self-contained exposition. Readers are encouraged to engage with it in the following ways:
\begin{itemize}
\item \textbf{Notation guide.}
Many notation and symbol conventions used in the VERaiPHY articles are established here. The glossary in Appendix~A provides the intended meaning. Where a symbol is overloaded across physics, statistics, and ML, the local context in each article will disambiguate.
 
\item \textbf{Conceptual primer.}
Physics readers may find the probability and statistics sections (\cref{sec:stats}) a useful entry point before engaging with articles that assume a frequentist or Bayesian inference background. Readers from the statistics or computer-science community may find the motivation in \cref{sec:ml} helpful for understanding where ML techniques have become central in fundamental physics. 
 
\item \textbf{Map to the series.}
Each VERaiPHY article addresses a specific methodological domain. Glossary entries marked \emph{(VERaiPHY article N)} indicate terms that receive a dedicated, in-depth treatment in a forthcoming contribution. Readers interested in a specific topic may navigate directly to the corresponding article, using this document to resolve any unfamiliar terminology encountered along the way.
\end{itemize}
Citations to original works, comparative discussions of methods, and pointers to the broader literature are provided in the dedicated VERaiPHY articles.
This document therefore cites only its companion contributions.
Its goal is to define concepts clearly and consistently rather than to survey the literature or to advocate for particular implementations.

Most importantly, VERaiPHY is an evolving project. 
Methods, terminology, and best practices at the intersection of ML and fundamental physics are constantly challenged by progress in the related fields and are developing much more rapidly than we are used to in pure physics. 
The conventions established here will be updated in response to community feedback or to align with emerging questions and concepts. 
Readers who identify missing terms, ambiguous definitions, or notation conflicts are encouraged to contact the authors.

\clearpage
\section{Foundations of Probability and Statistics}
\label{sec:stats}

This section introduces core probabilistic and statistical concepts that recur throughout the VERaiPHY series. It is not intended as a comprehensive textbook treatment; rather, it fixes terminology, notation, and interpretation to ensure consistency across contributions. Where multiple statistical interpretations exist (\eg frequentist and Bayesian), we briefly clarify their distinctions without advocating a particular philosophical stance.

\subsection{Basic Concepts}
\label{sec:basics}

We begin by recalling foundational definitions:

\begin{description}

\item[Probability (mathematical):]
Probability is defined axiomatically as a measure $P$ on a sample space $\Omega$ satisfying non-negativity, normalisation $P(\Omega)=1$, and countable additivity. These axioms provide a logically consistent foundation independent of interpretation.

\item[Probability (Bayesian):]
In the Bayesian interpretation, probability quantifies a degree of belief regarding the occurrence of an event or the value of an unknown parameter, such as a physical constant (\eg $m_\mathrm{W}$). Such probabilities may be assigned to single events or fixed but unknown constants and are updated via Bayes' theorem as new data becomes available. Since they reflect degrees of belief, such probabilities may differ between individuals with different prior knowledge or assumptions.

\item[Probability (frequentist):]
In the frequentist interpretation, probability is defined as the limiting relative frequency of an event in an infinite sequence of identical and independent trials. Probabilities are therefore not assigned to fixed but unknown parameters, only to outcomes of repeatable experiments.

\item[Conditional probability $p(A | B)$:]
The conditional probability $p(A | B)$ denotes the probability of $A$ given $B$, where $A$ and $B$ may represent events or hypotheses. In general, conditional probabilities are not symmetric, \ie $p(A | B) \neq p(B | A)$. Conditional probability forms the basis for statistical inference and for Bayes' theorem.

\item[Probability density:]
For discrete outcomes, probabilities are assigned to individual events, \eg $p_i = P(X = x_i)$. For continuous variables, probabilities are described by a probability density function (pdf) $p(x)$, such that probabilities are obtained by integration
\begin{equation}
P(m_1 \leq x \leq m_2) = \int_{m_1}^{m_2} \d x\, p(x)\;.
\end{equation}
The density itself need not be bounded by 1, but must integrate to unity over its domain.

\item[Bayes' theorem:]
Bayes' theorem relates conditional probabilities via
\begin{equation}
p(A | B) = \frac{p(B | A)\, p(A)}{p(B)}.
\end{equation}
This identity follows from the definition of conditional probability and is mathematically valid regardless of whether probability is interpreted in a frequentist or Bayesian sense. In statistical inference, one typically identifies $A$ with a parameter $\theta$ and $B$ with observed data $\mathcal{D}$, leading to
\begin{equation}
p(\theta | \mathcal{D}) = \frac{p(\mathcal{D} | \theta)\, p(\theta)}{p(\mathcal{D})}.
\end{equation}
Here $p(\mathcal{D} | \theta)$ is the likelihood, $p(\theta)$ the prior, $p(\theta | \mathcal{D})$ the posterior, and $p(\mathcal{D})$ the evidence (or marginal likelihood), which serves as a normalisation constant and plays a central role in model comparison and variational inference.

\item[Bayesian prior $p(\theta)$:]
The prior encodes information or assumptions about a parameter $\theta$ before observing the current data. Priors may be informative or weakly informative, depending on the context and available knowledge.

\item[Bayesian posterior $p(\theta | \mathcal{D})$:]
The posterior represents the updated probability distribution for $\theta$ after observing data $\mathcal{D}$. It combines prior information and the likelihood through Bayes' theorem.

\item[Marginalisation:]
If a probability distribution depends on multiple variables, the marginal distribution for a subset of them is obtained by integrating over the remaining variables
\begin{equation}
p(\theta) = \int \d\!\nu\;p(\theta, \nu)\;.
\label{eq:bayes_margin}
\end{equation}
Marginalisation is a fundamental operation in probability theory: it propagates uncertainty from unobserved or unwanted variables into the distribution of interest. 
A common application is the removal of nuisance
parameters $\nu$ from a joint posterior, but the same principle applies whenever one variable is to be eliminated. 

\item[Profiling:]
Given a function of multiple variables, its profile with respect to a subset of them is obtained by optimising over the remaining variables:
\begin{equation}
f_{\text{prof}}(\theta) = \max_{\nu}\, f(\theta, \nu)\;.
\label{eq:profiling}
\end{equation}
Profiling removes unwanted variables by replacing them with the values that optimise the function for each fixed $\theta$, thereby incorporating their
effect without integrating over them. 
Unlike marginalisation, profiling requires no probabilistic interpretation and applies to any function of multiple variables. 
The most common application is the profile likelihood, see Eq.\eqref{eq:profile_likelihood}. 
Profiling is the frequentist counterpart to marginalisation: where marginalisation averages over
uncertainty, profiling optimises it away.
The two approaches give identical results in Gaussian cases.

\item[Binomial distribution:]
For $n$ independent trials each with probability $\lambda$ of success, the probability of observing exactly $s$ successes is
\begin{equation}
p(s | n, \lambda) = \frac{n!}{s!(n-s)!}\lambda^s(1-\lambda)^{n-s}\;.
\end{equation}
Its mean is $n\lambda$ and its variance is $n\lambda(1-\lambda)$.

\item[Poisson distribution:]
The probability of observing $n$ events when the expected number is $\mu$ is given by
\begin{equation}
p(n | \mu) = e^{-\mu}\frac{\mu^n}{n!}\;.
\end{equation}
Both its mean and variance are equal to $\mu$. The Poisson distribution commonly arises as the limit of the binomial distribution for large $n$ and small $\lambda$ with fixed $n\lambda=\mu$.

\item[Gaussian (normal) distribution:]
The Gaussian (normal) distribution is defined as
\begin{equation}
\mathcal{N}(x | \mu,\sigma^2) = \frac{1}{\sqrt{2\pi}\sigma} 
\exp\left(-\frac{(x-\mu)^2}{2\sigma^2}\right)\;,
\end{equation}
with mean $\mu$ and variance $\sigma^2$.
The standard Gaussian corresponds to $\mu=0$ and $\sigma=1$.
The multivariate generalisation for a vector $x \in \mathbb{R}^d$ is
\begin{equation}
\mathcal{N}(x | \mu, \boldsymbol{\Sigma})
=
\frac{1}{(2\pi)^{d/2}|\boldsymbol{\Sigma}|^{1/2}}
\exp\left(-\frac{1}{2}(x-\mu)^\top \boldsymbol{\Sigma}^{-1}(x-\mu)\right)\;,
\end{equation}
where $\mu \in \mathbb{R}^d$ is the mean vector and $\boldsymbol{\Sigma}$ is the 
covariance matrix.
Gaussian approximations arise frequently due to the Central Limit Theorem and in large-sample limits of binomial, Poisson, and $\chi^2$ distributions.

\item[Student-$t$ distribution:]
The Student-$t$ distribution can be viewed as a Gaussian-like distribution with heavier tails, controlled by the number of degrees of freedom $\nu$. It is defined as
\begin{equation}
t(x| \mu,\sigma,\nu)
=
\frac{\Gamma\!\left(\frac{\nu+1}{2}\right)}
     {\Gamma\!\left(\frac{\nu}{2}\right)\sqrt{\nu\pi}\,\sigma}
\left[
1+\frac{1}{\nu}\left(\frac{x-\mu}{\sigma}\right)^2
\right]^{-\frac{\nu+1}{2}}\;,
\end{equation}
where $\mu$ is a location parameter, $\sigma$ a scale parameter, and $\nu$ the number of degrees of freedom. For large $\nu$, the distribution approaches a Gaussian. For $\nu=1$, it reduces to the Cauchy distribution, corresponding to the Lorentzian/Breit--Wigner form up to normalization conventions. 
A canonical example arises when the width of an approximately Gaussian distribution is estimated from the spread of $\nu+1$ observations rather than known exactly; the resulting distribution of the standardised variable then follows a Student-$t$ with $\nu$ degrees of freedom.
The Student-$t$ distribution is useful when one expects approximately Gaussian behaviour but with enhanced probability for large deviations.

\item[$\chi^2$ distribution:]
The $\chi^2$ distribution with $N$ degrees of freedom describes the distribution of the sum of squares of $N$ independent standard Gaussian variables. It is defined for $x\ge 0$ as
\begin{equation}
\chi_N^2(x)
=
\frac{1}{2^{N/2}\Gamma(N/2)}
x^{N/2-1} e^{-x/2}\;.
\end{equation}
Its mean is $N$ and its variance is $2N$. The $\chi^2$ distribution plays a central role in goodness-of-fit tests and in statistical constructions based on sums of squared residuals.

\item[Degrees of freedom:]
The number of degrees of freedom, often denoted $n_{\mathrm{dof}}$, counts the number of independent quantities entering a statistical construction after constraints or fitted parameters have been taken into account. For example, in a $\chi^2$ goodness-of-fit test with $N$ bins and $k$ fitted parameters, one often has approximately
\begin{equation}
n_{\mathrm{dof}} \approx N-k\;.
\end{equation}
The precise definition depends on the statistical procedure and on which quantities are treated as fixed or estimated from the data.

\item[Central Limit Theorem (CLT):]
The CLT states that the properly normalised sum of independent random variables with finite variance converges in distribution to a Gaussian as the number of terms increases. This result underlies the ubiquity of Gaussian approximations in statistical inference.

\end{description}

\subsection{Descriptive Statistics}
\label{sec:descriptive}

We next introduce basic statistical quantities used to summarise and characterise probability distributions.

\begin{description}

\item[Random variable:]
A random variable $x$ is a measurable function defined on a probability space, taking values in a specified domain and described by a probability distribution $p(x)$.

\item[Probability density function (pdf):]
A probability density function $p(x|\theta)$ describes the distribution of a continuous random variable $x$ under a model with parameters $\theta$. For fixed $\theta$, it satisfies
\begin{equation}
\int \d x \,p(x|\theta) = 1\;.
\end{equation}
Probabilities for intervals are obtained by integration over the corresponding domain.

\item[$p$-value:]
Given a hypothesis $H$, often the null hypothesis $H_0$, and a test statistic $T(\mathcal{D})$ computed from data $\mathcal{D}$, the $p$-value is the probability, under $H$, of observing a value of the test statistic at least as extreme as the one obtained:
\begin{equation}
p = P\big(T \ge T_{\mathrm{obs}} \,|\, H\big)\;,
\end{equation}
for a one-sided test (with appropriate modification for two-sided tests). Small $p$-values indicate tension between data and thehypothesis. Importantly, the $p$-value is \emph{not} the probability that $H$ is true. Under repeated experiments in which $H$ holds, $p$-values are uniformly distributed on $[0,1]$ for continuous test statistics; for discrete data the distribution is only approximately uniform, with $p$-values tending to be conservative.

\item[$z$-score:]
For any hypothesis test, a $p$-value may be expressed as an equivalent significance $z$, defined such that the one-sided Gaussian tail probability equals $p$:
\begin{equation}
z = \Phi^{-1}(1 - p)\;,
\end{equation}
where $\Phi$ is the standard normal CDF. This transformation is purely a re-expression of $p$ and does not require the test statistic itself to follow a Gaussian distribution. For example, a one-sided $z = 2$ corresponds to $p \approx 0.023$.

\item[Moments:]
For a random variable $x$ with density $p(x)$, the $n^\text{th}$ moment is defined as
\begin{equation}
\mathbb{E}[x^n] = \int \d x\, x^n\,p(x)\;.
\end{equation}

\item[Expectation:]
The expectation (mean) of $x$ is its first moment,
\begin{equation}
\mu \equiv \mathbb{E}[x] = \int \d x\, x\,p(x)\;.
\end{equation}

\item[Variance $\sigma^2$:]
The variance of $x$ measures the mean squared deviation from its expectation:
\begin{equation}
\sigma^2 = \mathbb{E}[(x-\mu)^2]
          = \int \d x\, (x-\mu)^2\,p(x)
          = \mathbb{E}[x^2] - \mu^2\;.
\end{equation}

\item[Bias:]
For an estimator $\hat\theta$ of a parameter $\theta$, the bias is defined as
\begin{equation}
\mathrm{Bias}(\hat\theta)
=
\mathbb{E}[\hat\theta]-\theta\;.
\end{equation}
An estimator is called unbiased if its expectation equals the true parameter value. Bias is a property of the estimator under repeated sampling and should be distinguished from the variance (the square of the uncertainty) of the estimates.

\item[Covariance $\mathrm{cov}(x_1,x_2)$:]
For random variables $x_1$ and $x_2$ with joint density $p(x_1,x_2)$, the covariance is
\begin{equation}
\mathrm{cov}(x_1,x_2) 
= \mathbb{E}[(x_1-\mu_1)(x_2-\mu_2)]
= \int \d x_1 \,\d x_2 \; (x_1-\mu_1)\,(x_2-\mu_2)\,p(x_1,x_2)\;.
\end{equation}
Covariance quantifies \emph{linear} dependence; it vanishes for uncorrelated variables.
For a vector of random variables $x=(x_1,\dots,x_n)$, the covariance matrix is defined by
\begin{equation}
\boldsymbol{\Sigma}_{ij}
=
\mathrm{cov}(x_i,x_j)\;.
\end{equation}
Its diagonal entries are the variances of the individual components, and the off-diagonal entries encode their pairwise covariances. 

\item[Correlation coefficient $\rho$:]
The Pearson correlation coefficient is defined as
\begin{equation}
\rho = \frac{\mathrm{cov}(x,y)}{\sigma_x \sigma_y}\;,
\end{equation}
and takes values in $[-1,1]$, with $\rho=\pm1$ corresponding to perfect linear correlation or anti-correlation.

\item[Independence:]
Random variables $x$ and $y$ are independent if their joint distribution factorises
\begin{equation}
p(x,y) = p_x(x)\,p_y(y)\;.
\end{equation}
Independence implies zero covariance (when moments exist), but the converse is not generally true.

\end{description}

\subsection{Distances and Divergences between Distributions}
\label{sec:distances}

In many applications one wishes to compare two probability distributions, for example a model prediction and a reference distribution, or two empirical samples.
These distributions may be defined over spaces of arbitrary dimension, and the measures discussed here apply in general to multi-dimensional data.
Several notions of discrepancy are commonly used for this purpose.
They differ in mathematical properties and in what aspects of the distributions they emphasise.

\begin{description}

\item[Statistical distance or divergence:]
A discrepancy measure quantifies the difference of two probability distributions $p$ and $q$.
Some of these are true metrics -- satisfying non-negativity, symmetry, and the triangle inequality -- while others are only divergences.
In the latter case, they still vanish when $p=q$, but need not be symmetric and need not satisfy the triangle inequality.

\item[$f$-divergence:]
A broad class of divergences between two distributions $p$ and $q$ is
given by the $f$-divergences,
\begin{equation}
D_f(p\|q)
=
\int \d x \, q(x)\,
f\!\left(\frac{p(x)}{q(x)}\right),
\end{equation}
where $f$ is a convex function with $f(1)=0$.
Different choices of $f$ define different divergences.
Examples include the Kullback--Leibler divergence, the reverse
Kullback--Leibler divergence, the Jensen--Shannon divergence, and the
Hellinger distance.
In general, $f$-divergences are not symmetric and are therefore not metrics.

\item[Kullback--Leibler divergence:]
The Kullback--Leibler (KL) divergence from $q$ to $p$ is defined as
\begin{equation}
D_{\mathrm{KL}}(p\|q)
=
\int \d x \, p(x)\,
\log\frac{p(x)}{q(x)}\;.
\end{equation}
It measures the expected logarithmic discrepancy between the two
distributions when samples are drawn from $p$.
It is non-negative and vanishes only if $p=q$ almost everywhere.
It is not symmetric:
\begin{equation}
D_{\mathrm{KL}}(p\|q)
\neq
D_{\mathrm{KL}}(q\|p)\;.
\end{equation}
The KL divergence plays a central role in information theory,
variational inference, and maximum-likelihood estimation.

\item[Jensen--Shannon divergence:]
A symmetrised and smoothed variant of the KL divergence is the
Jensen--Shannon divergence,
\begin{equation}
D_{\mathrm{JS}}(p,q)
=
\frac{1}{2}D_{\mathrm{KL}}(p\|m)
+
\frac{1}{2}D_{\mathrm{KL}}(q\|m),
\qquad
m = \frac{1}{2}(p+q)\;.
\end{equation}
Unlike the KL divergence, it is symmetric and always finite when
$p$ and $q$ are well-defined probability distributions.
Its square root defines a metric.

\item[Hellinger distance:]
The Hellinger distance between $p$ and $q$ is defined as
\begin{equation}
D_{\mathrm{H}}(p,q)
=
\frac{1}{\sqrt{2}}
\left(\int \d x\,\left(\sqrt{p(x)}-\sqrt{q(x)}\right)^2\right)^{1/2}\;,
\end{equation}
and satisfies $0 \le D_{\mathrm{H}}(p,q) \le 1$.
It is a true metric and is related to the total variation distance

\item[Total variation distance:]
The total variation distance between $p$ and $q$ is
\begin{equation}
D_{\mathrm{TV}}(p,q)
=
\frac{1}{2}
\int \d x \, |p(x)-q(x)|\;.
\end{equation}
It is a true metric and measures the difference in probability
assigned by the two distributions to the same event.

\item[Kolmogorov--Smirnov distance:]
The Kolmogorov--Smirnov (KS) distance between two univariate distributions $p$ and $q$ is the supremum of the pointwise difference between their cumulative distribution functions,
\begin{equation}
D_{\mathrm{KS}}(p,q)
=
\sup_{x}\,|F_p(x)-F_q(x)|\;,
\end{equation}
where $F_p$ and $F_q$ are the CDFs of $p$ and $q$ respectively.
It is a true metric, takes values in $[0,1]$, and is most sensitive to
differences near the centre of the distributions rather than in the tails.
It does not generalise straightforwardly to more than one dimension; see also the KS test in \cref{sec:gof}).

\item[Wasserstein distance:]
The Wasserstein distance compares two distributions by quantifying the
minimal cost of transporting probability mass from one distribution to
the other.
For example, the first Wasserstein distance is
\begin{equation}
W_1(p,q)
=
\inf_{\gamma \in \Gamma(p,q)}
\int \d x\,\d y \, \gamma(x,y)\,|x-y|\;,
\end{equation}
where $\Gamma(p,q)$ denotes the set of joint distributions $\gamma(x,y)$ with marginals
\begin{equation}
    p(x) = \int \d y\,\gamma(x,y) \qquad \mand \qquad
    q(y) = \int \d x\,\gamma(x,y)\;.
\end{equation}
Intuitively, $\gamma(x,y)$ specifies how much mass at location $x$ under $p$ is transported to location $y$ under $q$, and $|x-y|$ is the cost of that transport.
Unlike divergences based only on pointwise density ratios, Wasserstein distances are sensitive to the geometry of the underlying space.
They remain meaningful even when the supports of $p$ and $q$ do not overlap, and are therefore often useful for comparing structured or sample-based distributions.

\item[Choice of discrepancy measure:]
No single discrepancy measure is universally optimal.
KL-type divergences are often convenient when densities are available
and likelihood-based interpretations are desired.
Total variation provides a strict probabilistic notion of distinguishability.
Wasserstein distances are often preferable when geometric closeness of
samples matters, for example when comparing distributions over continuous
observables with nearby support.
The appropriate choice depends on the statistical task and on which
features of the distributions are most relevant.

\end{description}

\subsection{Modelling Physical Processes}
\label{sec:modelling}

We now introduce statistical concepts used when connecting probabilistic models to experimental data.

\begin{description}

\item[Parameter of interest $\theta$:]
Many analyses aim to estimate or constrain a parameter $\theta$, referred to as the parameter of interest. Additional parameters $\nu$, often associated with systematic uncertainties, are treated as nuisance parameters. In general, the likelihood depends on both $\theta$ and $\nu$.

\item[Likelihood:]
Given a model with parameter $\theta$ and probability density $p(\mathcal{D}|\theta)$, the likelihood function is the joint probability density of the observed dataset $\mathcal{D} = \{d_1,\dots,d_n\}$, viewed as a function of $\theta$:
\begin{equation}
\mathcal{L}(\theta)
= p(d_1,\dots,d_n | \theta).
\end{equation}
If the observations are statistically independent, this factorises as
\begin{equation}
\mathcal{L}(\theta)
= \prod_{i=1}^{n} p(d_i | \theta).
\end{equation}
The likelihood is not a probability density in $\theta$, since it does not integrate to one over $\theta$; rather, it is the data-generating density evaluated at the observed sample, considered as a function of the parameter. For histogrammed (binned) data, assuming independent Poisson fluctuations  in each bin, the likelihood becomes
\begin{equation}
\mathcal{L}(\theta) 
= \prod_{j} 
\frac{\lambda_j(\theta)^{\,b_j} e^{-\lambda_j(\theta)}}{b_j!}\;,
\end{equation}
where $b_j$ is the observed count in bin $j$ and $\lambda_j(\theta)$ is the 
predicted expectation value in bin $j$.

\item[Extended likelihood:]
If the total event yield itself depends on $\theta$, both the \emph{shape} and the \emph{normalisation} of the distribution carry information. In this case the likelihood includes the Poisson term for the total number of observed events, leading to the extended likelihood formulation:
\begin{equation}
\mathcal{L}_{\mathrm{ext}}(\theta)
=
e^{-\mu(\theta)}
\frac{\mu(\theta)^n}{n!}
\prod_{i=1}^{n} p(d_i|\theta)\;.
\end{equation}

\item[Profile likelihood:]
In the presence of nuisance parameters $\nu$, the profile likelihood is defined as
\begin{equation}
\mathcal{L}_{\text{prof}}(\theta) 
= \mathcal{L}\big(\theta, \hat{\nu}(\theta)\big)
\qquad \mwith \qquad
\hat{\nu}(\theta)
= \arg\max_{\nu} \mathcal{L}(\theta,\nu)\;,
\label{eq:profile_likelihood}
\end{equation}
where the nuisance parameters are replaced by the values that maximise the likelihood for fixed $\theta$. Profiling incorporates the effect of nuisance parameters by optimisation.  In Bayesian inference, nuisance parameters are instead eliminated by posterior marginalisation, as defined in Eq.\eqref{eq:bayes_margin}, which propagates uncertainty through averaging over their posterior distribution. A detailed discussion is provided in \emph{VERaiPHY article~4}~\cite{Dax:2026qsm}.

\item[Likelihood ratio:]
Likelihood-ratio statistics compare the relative support of the data for different parameter values or hypotheses. For two simple hypotheses, corresponding for example to parameter points $\theta_0$ and $\theta_1$, the likelihood ratio is
\begin{equation}
\Lambda(\mathcal D)
=
\frac{\mathcal{L}(\theta_1)}
     {\mathcal{L}(\theta_0)}\;.
\end{equation}
According to the Neyman--Pearson lemma, for testing two simple hypotheses this is the most powerful test statistic at fixed Type~I error rate.
In realistic applications, however, one typically deals with composite hypotheses and nuisance parameters. In that case, a commonly used test statistic for testing a fixed value of $\theta$ against the global best fit is the profile likelihood ratio,
\begin{equation}
\lambda(\theta)
=
\frac{\mathcal{L}(\theta,\hat\nu(\theta))}
     {\mathcal{L}(\hat\theta,\hat\nu)}\;,
\end{equation}
where $(\hat\theta,\hat\nu)$ denotes the global maximum-likelihood estimates. Likelihood-ratio statistics form the basis of many hypothesis tests and confidence interval constructions.

\item[Point estimation:]
A point estimate (best-fit) summarises the parameter of interest by a single value. In the frequentist approach, the maximum-likelihood estimate (MLE) is defined as
\begin{equation}
\hat{\theta}
= \arg\max_{\theta} \mathcal{L}(\theta)\;.
\end{equation}
In the Bayesian framework, a common choice is the maximum a posteriori (MAP) estimate,
\begin{equation}
\theta_\text{MAP}
= \arg\max_{\theta} p(\theta | \mathcal{D})\;,
\end{equation}
though other summaries such as the posterior mean or median may be used.

\end{description}

\subsection{Goodness of Fit \& Two-Sample Tests}
\label{sec:gof}

Goodness-of-fit and two-sample tests quantify the compatibility between observed data and a hypothesis $H$, which may contain free parameters determined from the data.
In a \emph{goodness-of-fit} test, one of the two distributions is known analytically -- either from theory or a parametric model fit to the data -- and the question is whether the observed sample is consistent with it. In a \emph{two-sample} test, both distributions are unknown and accessible only through finite samples; the question becomes whether the two samples are consistent with having been drawn from the same generative process. The two-sample setting is therefore strictly more general, and reduces to a goodness-of-fit problem when one sample size goes to infinity.
\begin{description}

\item[$\chi^2$ statistics:]
For binned data $d_i$ with theoretical expectations $t_i$,
a common test statistic is the weighted sum of squares
\begin{equation}
S = \sum_i \frac{(d_i - t_i)^2}{\sigma_i^2},
\end{equation}
or, more generally, in the presence of correlations,
\begin{equation}
S = \sum_{i,j}(d - t)_i^T \boldsymbol{\Sigma}_{ij}^{-1} (d - t)_j,
\end{equation}
where $\boldsymbol{\Sigma}$ is the covariance matrix.
Under suitable regularity conditions and for large samples,
$S$ follows approximately a $\chi^2$ distribution.

\item[Kolmogorov-Smirnov (KS) test:]
Given a sample $\{x_i\}_{i=1}^n$ and a reference CDF $F(x)$, the KS statistic measures the largest pointwise discrepancy between the empirical CDF $F_n(x)$ and the reference:
\begin{equation}
    D_n = \sup_{x} \left| F_n(x) - F(x) \right|
\end{equation}
The distribution of $D_n$ under $H$ is known analytically, provided $F$ is continuous, and is independent of $F$, so no simulation is needed to obtain
a $p$-value. The test is consistent -- it detects any deviation from $F$ given enough data -- but is most sensitive to differences near the median and loses power in the tails. It does not generalize to more than one dimension.

\item [Sliced tests.]
Sliced tests extend any one-dimensional test to multiple dimensions by projecting the data onto random unit vectors $\mathbf{v} \in \mathbb{S}^{d-1}$ and applying a 1D test to each projection:
\begin{equation}
    t_{\mathrm{sliced}} = \mathbb{E}_{\mathbf{v}}\left[ t\left( 
    \{\mathbf{v}^\top x_i\}_{i=1}^{n},\, 
    \{\mathbf{v}^\top y_j\}_{j=1}^{m} 
    \right) \right]
\end{equation}
where the expectation is estimated by averaging over $K$ random projections. Common instances include the Sliced Wasserstein distance and Sliced Kolmogorov-Smirnov test. The method is computationally attractive -- each projection reduces to a cheap 1D problem -- and scales well to high-dimensional data. However, sliced tests have limited sensitivity to distributional differences that manifest only in the \emph{joint} structure of the data, such as correlations: whether such differences are detected depends entirely on whether the sampled projections happen to be aligned with the directions of discrepancy.

\item[Maximum Mean Discrepancy (MMD).]
Given two distributions $p$ and $q$, the MMD measures their discrepancy as the distance between their mean embeddings in a reproducing kernel Hilbert space (RKHS) $\mathcal{H}$:
\begin{equation}
    \mathrm{MMD}^2(p, q;\, \mathcal{H}) = 
    \left\| \mathbb{E}_{x \sim p}[\phi(x)] - 
    \mathbb{E}_{y \sim q}[\phi(y)] \right\|^2_{\mathcal{H}}
\end{equation}
where $\phi: \mathcal{X} \to \mathcal{H}$ is the feature map associated to a kernel $k(x, x') = \langle \phi(x), \phi(x') \rangle_\mathcal{H}$. 
Applying the kernel trick, the squared MMD admits the unbiased estimator
\begin{equation}
    \widehat{\mathrm{MMD}}^2 = 
    \frac{1}{n(n-1)} \sum_{i \neq j} k(x_i, x_j)
    + \frac{1}{m(m-1)} \sum_{i \neq j} k(y_i, y_j)
    - \frac{2}{nm} \sum_{i,j} k(x_i, y_j)
\end{equation}
When $k$ is a \emph{characteristic} kernel, such as the Gaussian RBF
\begin{equation}
k(x,x') = \exp(-\|x - x'\|^2 / 2\sigma^2)\;,
\end{equation}
the MMD is a proper metric on probability distributions: $\mathrm{MMD}(p, q) = 0$ if and only if $p = q$. 
This means the test is sensitive to \emph{all} moments of the distributions, including correlations and higher-order structure, unlike sliced tests or marginal comparisons. 
In practice, the estimator variance grows with dimensionality, so MMD works best in low-to-moderate dimensions or after a suitable dimensionality reduction step; extensions based on learned kernels or Nystr\"{o}m approximations have been proposed to improve scalability to large datasets.

\item[Pulls:]
A pull is a normalized residual used to quantify the local discrepancy between data and prediction. A common definition is
\begin{equation}
r_i
=
\frac{d_i-t_i}{\sigma_i}\;,
\end{equation}
where $d_i$ is the observed value, $t_i$ the corresponding prediction, and $\sigma_i$ the uncertainty on the prediction $t_i$.
Pull distributions are useful diagnostic tools for identifying local mismodelling, underestimated uncertainties, or systematic trends across bins or observables.

\end{description}
It is worth noting that any divergence or distance between probability distributions -- such as the Kullback-Leibler divergence, the 
Jensen-Shannon divergence, or the Wasserstein distance described in \cref{sec:distances} -- can in 
principle be used as a two-sample test statistic, provided it can be estimated from finite samples. 
The key practical distinction is that divergences are defined between distributions, whereas in the two-sample setting both $p_X$ and $p_Y$ are unknown and must be estimated from data; the quality of the test 
therefore depends critically on how well the divergence estimator behaves in finite-sample, high-dimensional regimes. In this sense, the MMD can itself be understood as an instance of this broader strategy: it is a distance between distributions that happens to admit a simple, unbiased, closed-form estimator from samples.

A detailed discussion of goodness-of-fit diagnostics,
model misspecification, and high-dimensional extensions
is deferred to in \emph{VERaiPHY article~3}~\cite{Cruz-Martinez:2026apu} and \emph{VERaiPHY article~6}~\cite{Amram:2026vkc}.

\subsection{Hypothesis Testing}
\label{sec:hypo}

Hypothesis testing assesses which of two competing hypotheses,
$H_0$ and $H_1$, is better supported by the data. Unlike parameter estimation, hypothesis testing is a discrete problem.
In particle physics this often corresponds to comparing
the Standard Model with an extension that includes
a specific form of new physics.

\begin{description}

\item[Test statistic:]
A test statistic $T(\mathcal{D})$ is a low-dimensional summary of the data
used in the inference procedure. For simple counting experiments, $T$ may be the observed number of events. In more complex analyses, a common choice is the likelihood ratio introduced earlier. For testing two simple hypotheses, the Neyman--Pearson lemma states that the likelihood-ratio test is the most powerful test at fixed Type~I error rate.

\item[Nested hypotheses:]
Two hypotheses are nested if the larger can be reduced
to the smaller by fixing particular parameter values.
For nested models and under suitable regularity conditions,
likelihood-ratio statistics obey asymptotic results.

\item[Asymptotics:]
Many statistical results used in inference are asymptotic, meaning that they become valid in the limit of large sample size. Asymptotic formulae are often extremely useful in practice, but their accuracy for finite samples depends on the problem and on the validity of the required regularity conditions.

\item[Wilks' theorem:]
For nested hypotheses and regular models,
\begin{equation}
-2\log\lambda(\theta)
\;\sim\;
\chi^2_k
\qquad \text{for } n\to\infty
\end{equation}
with $k$ equal to the difference in parameter dimensionality between the models.
Important regularity conditions must be satisfied for this result to hold.

\item[Discovery:]
In particle physics, a discovery is conventionally claimed when the local $p$-value for the null hypothesis corresponds to a significance of $5\sigma$ (approximately $3\times10^{-7}$).

\item[Look-elsewhere effect:]
When scanning over multiple regions of parameter space, a distinction must be made between the local $p$-value
(at a fixed parameter point) and the global $p$-value, which accounts for the multiplicity of tests.
The definition of the search domain must therefore be specified.

\item[Limits:]
Upper limits on the signal strength are typically reported at $95\%$ confidence or credible level. 
Such limits may translate into lower bounds on masses or couplings.
The choice of interval type -- whether a two-sided interval or a one-sided upper limit -- should be made prior to examining the data; deciding based on the outcome (\eg reporting 
an upper limit only when no excess is observed) is known as flip-flopping and leads to intervals with incorrect coverage properties.

\item[$\mathrm{CL}_\mathrm{s}$ method:]
In exclusion analyses, one often uses the modified-frequentist $\mathrm{CL}_\mathrm{s}$ prescription to avoid excluding models to which the experiment has little sensitivity. It is commonly defined as
\begin{equation}
\mathrm{CL}_\mathrm{s}
=
\frac{p_{s+b}}{1-p_b}\;,
\end{equation}
where $p_{b}$ and $p_{s+b}$ denote the relevant $p$-values under the background-only and signal-plus-background hypotheses, respectively. Exclusion is then typically declared when $\mathrm{CL}_\mathrm{s}$ falls below a chosen threshold, such as $0.05$.

\item[Type I and Type II errors:]
A Type~I error corresponds to rejecting a true null hypothesis,
while a Type~II error corresponds to failing to reject a false one.
In classification problems (\eg signal versus background),
these errors correspond to the background misidentification rate and to one minus the signal efficiency, respectively.

\item[ROC curves:]
Receiver Operating Characteristic (ROC) curves display classifier performance as the decision threshold is varied.
In the statistics and ML convention one plots the true positive rate against the false positive rate, while in high-energy physics the same curve is usually drawn as the background misidentification rate, or its inverse the background rejection, against the signal efficiency.
The two terms are equivalent: signal efficiency corresponds to the true positive rate, and background misidentification rate to the false positive rate.
See \cref{sec:supervised}.

\end{description}
A more detailed discussion of hypothesis testing procedures, asymptotic formulae, calibration, multiple-testing effects, and their connections to modern anomaly-detection strategies is provided in \emph{VERaiPHY article~6}~\cite{Amram:2026vkc}.

\subsection{Intervals for estimated parameters}
\label{sec:intervals}

Intervals provide more informative summaries than point estimates,
as they quantify the uncertainty associated with the parameter of
interest.

\begin{description}

\item[Likelihood-based intervals:]
In the frequentist approach, confidence intervals can be constructed from the likelihood function. For a single parameter $\theta$, a common large-sample approximation is to expand $\log \mathcal{L}(\theta)$ around its maximum at $\hat\theta$. If the log-likelihood is approximately quadratic, the $1\sigma$ interval is given by the region where
\begin{equation}
\log \mathcal{L}(\hat\theta)-\log \mathcal{L}(\theta)=\frac{1}{2}\;,
\end{equation}
equivalently $-2\Delta\log\mathcal{L}=1$. The same approximation leads to the curvature estimate
\begin{equation}
\sigma_\theta^2
=
\left(
- \frac{\partial^2 \log \mathcal{L}}{\partial \theta^2}
\right)^{-1}_{\theta=\hat\theta}.
\end{equation}
These constructions rely on the likelihood being sufficiently regular and approximately Gaussian near its maximum.

\item[Bayesian credible intervals:]
In Bayesian inference, credible intervals are derived from the posterior distribution $p(\theta | \mathcal{D})$. 
One may construct central intervals, highest-posterior-density (shortest) intervals, or one-sided limits, all at a specified credible level. These intervals are referred to as credible intervals. 
Bayesian credible intervals depend on the choice of parametrisation; the posterior $p(\theta|\mathcal{D})$ changes shape under a nonlinear reparametrisation, so an interval constructed for $\theta$ will not in general correspond to the transformed interval for $f(\theta)$.

\item[Asymmetric uncertainties:]
Uncertainty intervals need not be symmetric around a central estimate. This commonly occurs when the likelihood or posterior is skewed, when physical boundaries are present, or when the parameter dependence is non-linear. In such cases one often quotes results in the form
\begin{equation}
\hat\theta^{+\,\Delta\theta^+}_{-\,\Delta\theta_-}\;,
\end{equation}
with separate upper and lower uncertainties.

\item[Neyman construction:]
A frequentist procedure for constructing confidence intervals for a parameter $\theta$ based on the sampling distribution $p(\mathcal{D} | \theta)$. When the underlying assumptions are satisfied, the resulting intervals achieve coverage at or above the nominal confidence level in repeated experiments.

\item[Feldman--Cousins construction:]
Confidence intervals can be constructed using the Neyman construction. Within that framework, the choice of ordering rule determines whether the resulting interval is central, one-sided, or of Feldman--Cousins type. The Feldman--Cousins construction uses an ordering principle based on likelihood ratios and provides a unified frequentist procedure that automatically interpolates between two-sided confidence intervals and one-sided upper limits, thereby avoiding ad hoc switching between interval types.

\item[Coverage:]
For a procedure that reports intervals at confidence level $1-\delta$, the coverage $\mathcal{C}(\theta)$ is the probability that the interval contains the true value of $\theta$, under repeated experiments. 
Coverage is therefore a criterion for assessing whether reported intervals are statistically well calibrated.
Coverage is a property of the statistical procedure itself, not of a single realised interval. 
In the presence of nuisance parameters and conditioning
variables, one may distinguish between conditional and marginal
coverage. 
A detailed discussion of coverage properties, calibration, and their relation to inference procedures is deferred to \emph{VERaiPHY article~2}~\cite{Haussmann:2026gbi} and \emph{VERaiPHY article~4}~\cite{Dax:2026qsm}.

\item[Expected results:]
In addition to reporting intervals obtained from observed data, it is often useful to quote expected results under an assumed hypothesis. A common example is the expected interval or expected upper limit obtained from pseudo-experiments or from an Asimov dataset (\ie a dataset equal to its expectation). Bands corresponding to typical fluctuations (e.g.\ $68\%$, $95\%$) are frequently shown. 
Expected results are also commonly required when assessing the 
projected sensitivity of future experiments, for instance in 
proposals for funding or detector upgrades.

\item[Uncertainties:]
Uncertainties are commonly categorised as statistical or systematic. Statistical uncertainties arise from finite sample size and typically scale as $1/\sqrt{N}$, where $N$ is the number of observations, under repeated sampling.
Systematic uncertainties originate from imperfect modelling, instrumental effects, theoretical assumptions, or external inputs, and do not necessarily decrease with additional data. 
In particle physics, systematic uncertainties are typically described by nuisance parameters, which can be Gaussian.
A detailed discussion of uncertainty modelling, propagation, correlations, and coverage properties is beyond the scope of this introductory section and is addressed in a dedicated article of the VERaiPHY series.
Theory uncertainties affect many measurements in particle physics and can be viewed as systematic uncertainties, but the distribution of the corresponding nuisance parameters defy statistical interpretation. 

\end{description}

\subsection{Combining Results}

Combining results from different analyses requires careful treatment of statistical and systematic correlations. In modern applications, several measurements often constrain several parameters simultaneously, possibly in the presence of shared nuisance parameters. For this reason, it is generally preferable to combine analyses at the likelihood level rather than combining only final quoted estimates.

Suppose a set of measurements or analyses, labeled by $a$, constrains a common parameter vector $\theta$. A combined analysis is then naturally formulated through a joint likelihood
\begin{equation}
\mathcal{L}_{\mathrm{comb}}(\theta,\nu)
=
\prod_a \mathcal{L}_a(\theta,\nu_a,\nu_{\mathrm{corr}})\;,
\end{equation}
where $\nu_a$ denotes nuisance parameters specific to analysis $a$, and $\nu_{\mathrm{corr}}$ denotes nuisance parameters shared across several analyses. This formulation allows to propagate correlated systematic uncertainties consistently, account for parameter degeneracies, and retain non-Gaussian information that may be lost when only final estimates are combined.

Inferences are then based on the combined likelihood in the same way as for a single analysis, for example through maximum-likelihood estimation, profile likelihood ratios, confidence intervals, or Bayesian posterior distributions. In this sense, simple weighted averages should be regarded as a special limiting case of a more general statistical combination problem.

\begin{description}

\item[Linear combinations:]
When several independent measurements $x_i \pm \sigma_i$ constrain the same scalar quantity $x$, and the uncertainties are approximately Gaussian, the minimum-variance linear estimator is
\begin{equation}
\hat{x} = \sum_i w_i x_i
\qquad \mwith \qquad
w_i = \frac{1/\sigma_i^2}{\sum_j 1/\sigma_j^2}\;,
\end{equation}
with variance
\begin{equation}
\sigma^{-2} = \sum_j \frac{1}{\sigma_j^2}\;.
\end{equation}
Thus, measurements with smaller uncertainties receive larger weight.
In the presence of correlations, the optimal linear combination is obtained from the inverse covariance matrix,
\begin{equation}
\hat{x}
= \frac{\sum_{i,j} (\boldsymbol{\Sigma}^{-1})_{ij}\, x_j}
     {\sum_{i,j} (\boldsymbol{\Sigma}^{-1})_{ij}}\;,
\end{equation}
with corresponding variance
\begin{equation}
\sigma^2
=
\left(
\sum_{i,j} (\boldsymbol{\Sigma}^{-1})_{ij}
\right)^{-1}\;.
\end{equation}
In such cases, the effective weights can lie outside the interval $[0,1]$, reflecting the structure of correlations rather than any inconsistency of the procedure.

\item[Multi-parameter combinations:]
More generally, several measurements may constrain several parameters simultaneously. In such cases, the combination cannot be reduced to a single weighted average. Instead, one performs a joint inference for the full parameter vector, taking into account both parameter correlations and shared nuisance parameters. This is particularly important when different measurements constrain complementary directions in parameter space, or when systematic effects induce substantial correlations across analyses.

In approximately Gaussian problems, this generalization can be expressed in terms of a vector of observables and a covariance matrix. More generally, however, the full likelihood formulation is preferred, since it naturally accommodates non-linear parameter dependences, non-Gaussian uncertainties, limited data, and information about nuisance parameters; see also the discussion of combining test results below. 

\item[Combining test results:]
Combining $p$-values, significances, or hypothesis-test outcomes requires particular care. 
There is no unique method for combining $p$-values, and 
two small $p$-values may reflect discrepancies in different aspects of the data or different regions of parameter space, so their combination can obscure rather than clarify the underlying tension. 
More generally, such quantities are not, in general, sufficient statistics for a full combination, especially in the presence of correlations, nuisance parameters, or multiple testing. 
Whenever possible, hypothesis tests should therefore be recomputed from a joint likelihood or from the underlying data model rather than from already reduced summaries.

\end{description}

\subsection{Sampling and Resampling}
\label{sec:sampling}

Sampling and resampling methods form a core toolkit in statistical inference. 
They enable the approximation of expectations, the estimation of uncertainties, hypothesis testing, and model validation -- particularly in settings where analytic solutions are unavailable.
These techniques are central to both frequentist and Bayesian frameworks and play a crucial role in simulation-based workflows common in particle physics.

\begin{description}

\item[Monte Carlo integration:]
Monte Carlo (MC) integration estimates integrals by averaging function evaluations at randomly sampled points.
In its simplest form (naive Monte Carlo), one samples uniformly over the integration domain.
For a function $f(x)$ defined on a domain $\Omega$,
\begin{equation}
\int_\Omega \d x\, f(x)
\approx
\frac{V}{N} \sum_{i=1}^{N} f(x_i),
\qquad x_i \sim \text{uniform}(\Omega)\;,
\end{equation}
where $V$ denotes the volume of $\Omega$.
More generally, one may sample from a probability density $p(x)$ and rewrite
\begin{equation}
\int_\Omega \d x\, f(x)
= \int \d x\, p(x)\,\frac{f(x)}{p(x)}
\approx
\frac{1}{N} \sum_{i=1}^{N}
\frac{f(x_i)}{p(x_i)},
\qquad x_i \sim p(x)\;.
\end{equation}
If $p(x)$ is chosen to approximate the shape of $f(x)$, \ie $p(x) \approx C\, f(x)$, the variance of the estimator can be significantly reduced. This strategy is known as \emph{importance sampling} and is widely used in physics, for example in the evaluation of high-dimensional phase-space integrals.
A key advantage of standard Monte Carlo integration is that its statistical error scales as $\mathcal{O}(1/\sqrt{N})$, independently of dimensionality, which makes it particularly robust for high-dimensional problems. Other sampling strategies, such as quasi-Monte Carlo (QMC), which use structured, non-i.i.d.\ and effectively correlated sample sets, can in favourable cases exhibit faster convergence.

\item[Monte Carlo sampling:]
Monte Carlo sampling refers to the generation of pseudo-random samples from a specified probability distribution in order to approximate expectations, propagate uncertainties, or study the distribution of derived quantities. Beyond its use for numerical integration, it provides a general tool for inference and uncertainty propagation: if $z^{(k)} \sim p(z)$, then expectations of a function $f(z)$ can be approximated through sample averages,
\begin{equation}
\mathbb{E}[f(z)] \approx \frac{1}{N}\sum_{k=1}^N f(z^{(k)})\;.
\end{equation}
In statistical applications, this idea is often used to propagate uncertainty from model parameters to observables. In a Bayesian setting, for example, one may draw samples from the posterior distribution -- or, in iterative schemes such as Gibbs sampling, from conditional posterior distributions -- and evaluate the quantity of interest for each draw, thereby obtaining both point estimates and uncertainty intervals for derived observables.

\item[Markov Chain Monte Carlo:]
Markov Chain Monte Carlo (MCMC) methods generate samples from a target probability distribution by constructing a Markov chain whose stationary distribution equals the target. 
Starting from an initial state, the chain evolves by proposing and accepting or rejecting new states according to an acceptance criterion that ensures convergence 
to the target. 
MCMC is widely used in Bayesian inference to sample 
from intractable posterior distributions, and forms the basis of algorithms such as the Metropolis--Hastings algorithm and Hamiltonian Monte Carlo.

\item[Bootstrap:]
The bootstrap is a resampling-based method for estimating the variability of a statistic, such as its variance or confidence interval. Given a dataset of $n$ observations, one draws $B$ bootstrap samples of size $n$ with replacement and evaluates the statistic of interest for each sample.
The resulting distribution of bootstrap estimates approximates the sampling distribution of the statistic.
Bootstrap methods are widely used in astrophysics and cosmology, for example to estimate uncertainties on power spectra or light-curve fits without assuming a parametric form for the underlying data distribution.

\item[Permutation tests:]
Permutation (or randomisation) tests assess statistical significance by constructing the null distribution of a test statistic through random relabelling of the data.
For example, to test whether two samples $A$ and $B$ have different means, one computes the observed mean difference and compares it to the distribution obtained by repeatedly shuffling the sample labels.
This approach relies on exchangeability under the null hypothesis and makes minimal distributional assumptions.
It is particularly useful when analytic null distributions are unavailable but symmetry arguments justify label permutation.

\item[Cross-validation:]
Cross-validation evaluates the predictive performance and generalisation error of a statistical model.
In $k$-fold cross-validation, the dataset is partitioned into $k$ subsets; the model is trained on $k-1$ subsets and evaluated on the remaining subset. This procedure is repeated $k$ times, and the cross-validation estimate is
\begin{equation}
\text{CV}_k
=
\frac{1}{k}
\sum_{i=1}^{k}
L_{\text{test}}^{(i)},
\end{equation}
where $L_{\text{test}}^{(i)}$ denotes the loss on fold $i$.
Cross-validation is particularly valuable when data is limited, as it provides a more stable estimate of predictive performance than a single train/test split and helps diagnose overfitting.

\item[Toy Monte Carlo:]
Toy Monte Carlo, often also called pseudo-experiments, refers to the repeated generation of synthetic datasets from a statistical model in order to study the behaviour of estimators, test statistics, confidence intervals, or full analysis procedures under repeated sampling. Such simulations are widely used to estimate expected sensitivities, assess biases, validate coverage, and determine the sampling distribution of test statistics. In realistic applications, nuisance parameters may also be varied or sampled according to their constraint model in order to propagate systematic uncertainties and study their impact on the final inference.
Related Monte Carlo ideas also appear in Bayesian inference, where samples from posterior distributions are used to propagate parameter uncertainties to derived observables.

\end{description}
Modern ML-based generative models offer complementary approaches to simulation and data augmentation, including generative amplification; these are discussed 
in \cref{sec:generative} and in \emph{VERaiPHY article~5}~\cite{Diefenbacher:2026kki}.

\clearpage
\section{Foundations of Machine Learning}
\label{sec:ml}

Machine learning can be viewed as a modern extension of statistical modelling, augmented with flexible function approximators and scalable optimization techniques. It provides a framework for learning predictive relationships directly from data, particularly in settings where explicit analytical modelling is difficult or infeasible. 
At a conceptual level, most machine-learning workflows involve three ingredients: a \emph{learning objective}, a \emph{model class} used to represent the mapping from inputs to outputs, and a \emph{training procedure} that adjusts the model parameters by minimizing a suitable loss function. From a physicist's perspective, this process is closely related to statistical estimation and variational optimisation, but typically places greater emphasis on flexible data-driven representations.

\subsection{Why Machine Learning in Fundamental Physics?}

The growing use of machine learning (ML) in fundamental physics is driven by both practical need and scientific opportunity. Modern experiments and simulations routinely produce high-dimensional, heterogeneous, and large-scale datasets, ranging from raw detector signals in collider experiments to simulated fields, maps, or time series in cosmology and astrophysics. In many such settings, conventional low-dimensional summaries or analytically tractable likelihoods are unavailable, insufficient, or computationally too expensive.

Machine learning provides a flexible set of tools for dealing with such problems. ML models can approximate highly non-linear mappings, extract relevant features from complex data, and accelerate computationally expensive parts of simulation or inference pipelines. Typical applications include classification, regression, anomaly detection, reconstruction, surrogate modelling, and fast simulation.

Beyond computational efficiency, ML also opens new paths for scientific analysis. It enables inference in implicit models, provides flexible function classes for inverse problems, and supports differentiable pipelines in which simulation, optimisation, and statistical inference can be combined. In this sense, ML is not only a computational tool, but also a modelling framework that complements first-principles reasoning with data-driven adaptability.

\subsection{Core Concepts of Statistical Learning}
\label{sec:coreml}

We next introduce the basic concepts underlying supervised and unsupervised ML, many of which are straightforward generalizations from fitting the parameters of a given function to data.

\begin{description}

\item[Model:]
A model is a function $f$ or probability distribution $p(y|x)$ used to represent the relationship between inputs and outputs. The model may be \textit{parametric}, characterized by a fixed parameter vector $\theta$ learned from data, or \textit{non-parametric}, where the model complexity adapts to the data. Common model classes include linear models, kernel methods, decision trees, and neural networks.

\item[Learning task:]
A learning task specifies which statistical relation is to be learned from data.
In \textit{supervised learning}, one seeks a function mapping inputs to outputs,
\begin{equation}
f:\mathcal{X}\to\mathcal{Y} \; ,
\end{equation}
where the model is trained on labeled pairs $(x, y)$. In fundamental physics, 
the feature space $\mathcal{X}$ is often some kind of phase space. 
Common supervised tasks include regression, where $\mathcal{Y}$ is continuous 
and the output is a conditional distribution $p(y|x)$ -- possibly a deterministic 
prediction $\delta(y - f(x))$ as a special case -- and classification, where 
$\mathcal{Y}$ is discrete and $p(y|x)$ can equivalently be expressed as a 
likelihood ratio $\frac{p(x|\theta_1)}{p(x|\theta_0)}$.

In \textit{unsupervised learning}, no target labels are provided, and the goal 
is instead to learn the structure of the data itself. Generative models, for 
instance, learn a distribution $p(x)$ or a conditional distribution $p(x|c)$, 
where $c$ is an optional condition. More details on supervised and unsupervised 
learning follow below.

\item[Dataset:]
A dataset is a finite collection of examples,
\begin{equation}
\mathcal{D} = \{(x_i,y_i)\}_{i=1}^N
\end{equation}
in the supervised case, or $\mathcal{D}=\{x_i\}_{i=1}^N$ in the unsupervised case. It is commonly split into training, validation, and test subsets.
The validation set is used to guide model selection, hyperparameter tuning, or early stopping. The test set is reserved for the final evaluation of model performance and should not be used during training or model design.

\item[Parameters and hyperparameters:]
In parametric models, the parameters $\theta$ are adjusted during training. 
By contrast, hyperparameters are external choices that specify aspects of the 
model or training procedure, such as network depth, learning rate, regularization 
strength, or batch size. These hyperparameters lead to significant inductive bias, 
for example the resolution of a network function over feature space.

\item[Loss function:]
The loss function $L(\theta)$ quantifies the discrepancy between model predictions 
and training targets. The minimization of a loss function can be derived from a 
maximum a posteriori (MAP) estimate of the optimal parameters $\theta$ given the 
training data,
\begin{align}
  L(\theta) 
  &= - \log p(\theta|\mathcal{D}) \notag \\
  &= - \log p(\mathcal{D}|\theta) \;
    - \log p(\theta) + \text{const}(\theta) \; .
\end{align}
The likelihood $p(\mathcal{D}|\theta)$ is often assumed to be Gaussian, motivated 
for instance by the central limit theorem, but this is not a fundamental requirement 
and other likelihood models may be more appropriate depending on the data. The prior 
$p(\theta)$ encodes beliefs about the parameters before observing data and acts as 
a regulariser. 
In the special case of a uniform prior, MAP reduces to maximum likelihood estimation (MLE).
Training then refers to the optimisation procedure used to adjust the model 
parameters. In modern ML, this is usually carried out by stochastic gradient-based 
methods applied to the empirical loss on the training data.

\item[Generalization:]
A central goal in ML is not merely to fit the training data, but to perform well on previously unseen data drawn from the same or a related distribution. This ability is referred to as generalization, and it is achieved through explicitly or implicitly chosen inductive bias.
A detailed discussion is provided in \emph{VERaiPHY article~1}~\cite{Lavie:2026psw}.

\item[Regularization:]
Regularization improves generalization by constraining the flexibility of the model 
or by stabilizing training, and can be understood naturally in the MAP framework 
introduced above. The prior $p(\theta)$ encodes beliefs about the parameters before 
observing data, and $-\log p(\theta)$ acts as a regularization term in the loss. 
Concretely, a Gaussian prior
\begin{equation}
    p(\theta) \propto e^{-\lambda\|\theta\|^2}
\end{equation}
corresponds to L2 regularization (weight decay), while a Laplace prior corresponds to L1 
regularization, promoting sparsity. More sophisticated priors, such as those arising 
in Bayesian neural networks, can be shown to correspond to dropout regularization.

Beyond explicit prior-based penalties, regularization can also be applied ad-hoc 
through other means, including early stopping, data augmentation, and architectural 
choices such as network depth or width. Regularization is used to avoid overfitting, 
where a model adapts to statistical fluctuations or accidental structure in the 
training dataset and consequently degrades its generalization to unseen test data.

\item[Representation learning:]
Representation learning refers to methods that automatically learn useful features or internal representations of the data, rather than relying on hand-crafted features. 
The goal is to find a representation that captures the underlying structure of the data in a way that makes downstream tasks such as classification or regression easier. Common 
approaches include autoencoders, contrastive learning, and 
self-supervised methods. In physics, representation learning is 
relevant for learning compact descriptions of complex events or 
fields that preserve physically meaningful structure.
A detailed discussion is given in \emph{VERaiPHY article~9}~\cite{VERaiPHY_rep}.

\item[Equivariance and invariance:]
Given a symmetry operation $S$, a function can be designed to respect that symmetry either as equivariance, $f(S(x)) = S(f(x))$, meaning the output transforms in the same way as the input, or as invariance, $f(S(x)) = f(x)$, meaning the output is unaffected by the symmetry operation. Incorporating known symmetries into a model is a powerful form of inductive bias, particularly relevant in physics where symmetries play a fundamental role. 
A detailed discussion is given in \emph{VERaiPHY article~7}~\cite{VERaiPHY_sym}.

\end{description}

\subsection{Gradient-Based Optimisation}

Most modern ML models, including neural networks, are trained by minimizing a loss function $L(\theta)$ with respect to the model parameters $\theta$. In practice, this optimisation is usually performed by iterative first-order methods based on gradients.

\begin{description}

\item[Gradient descent:]
The simplest gradient-based update takes the form
\begin{equation}
\theta^{(t+1)} =
\theta^{(t)} - \eta \nabla_\theta L(\theta^{(t)})\;,
\end{equation}
where $\eta$ is the learning rate. The update follows the direction of steepest local decrease of the loss.

\item[Stochastic gradient descent:]
For large datasets, the gradient is typically estimated using a randomly selected mini-batch $\mathcal{B}\subset\mathcal{D}$ rather than the full dataset. This leads to the stochastic gradient descent (SGD) update
\begin{equation}
\theta^{(t+1)} =
\theta^{(t)} - \eta \nabla_\theta L_{\mathcal{B}}(\theta^{(t)})\;.
\end{equation}
The stochasticity of the mini-batch estimate reduces computational cost and can help the optimisation escape shallow local structures.

\item[Momentum:]
Momentum methods introduce an auxiliary velocity variable $v$ that accumulates information from previous gradients:
\begin{align}
v^{(t+1)} &= \gamma v^{(t)} - \eta \nabla_\theta L(\theta^{(t)}), \notag\\
\theta^{(t+1)} &= \theta^{(t)} + v^{(t+1)}\;,
\end{align}
where $\gamma\in[0,1)$ is the momentum coefficient. This can accelerate convergence, especially in ill-conditioned loss landscapes.

\item[Adaptive optimizers:]
Adaptive methods adjust the effective learning rate separately for different parameters based on the history of past gradients. Widely used examples include RMSprop and Adam. Such optimizers often improve practical training stability, especially for deep or noisy models, although their effect on generalization can depend strongly on the application.
Adam (Adaptive Moment Estimation), for instance, combines momentum-like first-moment estimates with adaptive second-moment scaling. Denoting by $m^{(t)}$ and $v^{(t)}$ the running estimates of the first and second gradient moments, the bias-corrected update is
\begin{align}
\hat{m}^{(t)} &= \frac{m^{(t)}}{1-\beta_1^t}, \qquad
\hat{v}^{(t)} = \frac{v^{(t)}}{1-\beta_2^t}, \notag\\
\theta^{(t+1)} &= \theta^{(t)} - \frac{\eta}{\sqrt{\hat v^{(t)}}+\epsilon}\,\hat m^{(t)}\;.
\end{align}
Adam is widely used because it is robust and easy to tune in many applications.

\item[Learning-rate schedules:]
In many training procedures the learning rate is varied during optimization rather than kept fixed. Common choices include step decay, exponential decay, cosine annealing, and warm restarts. Such schedules often improve convergence and can help refine the solution during later stages of training.

\end{description}
In physics-inspired applications, the choice of optimiser can substantially affect training stability, computational efficiency, and generalization performance, especially in problems with stiff constraints, noisy gradients, or highly multi-scale loss landscapes.

\subsection{Supervised Learning: Classification and Regression}
\label{sec:supervised}

Supervised learning refers to the setting in which a model is trained on 
labelled examples in order to predict targets from inputs. The two most 
common cases are regression, where the target $y$ is continuous, and 
classification, where the target belongs to a finite set of discrete classes. 
In both cases, the model learns a mapping $f: \mathcal{X} \to \mathcal{Y}$ 
from labeled training pairs $(x_i, y_i)$, as introduced in \cref{sec:coreml}.

\begin{description}

\item[Regression:]
In regression tasks, the target variable is a continuous quantity, typically a scalar. 
Examples in physics include energy calibration, parameter reconstruction, emulator construction, transition probabilities, or scattering amplitudes.

\item[Classification:]
In classification tasks, the target belongs to a finite set of classes. 
Common examples include signal-versus-background discrimination, event categorisation, source identification, and morphology classification. 
The network output is a probability distribution over the set of classes $\mathcal{Y}$, representing the predicted probability that a given input $x$ belongs to each class.

\item[Model classes:]
Supervised learning can be carried out with many different model families, including linear and logistic regression, decision trees, support vector machines, kernel methods, and neural networks. These models differ in expressiveness, inductive bias, interpretability, and computational cost.

\item[Mean squared error:]
A standard loss for regression is the mean squared error (MSE),
\begin{equation}
L_{\mathrm{MSE}}
=
\frac{1}{n}\sum_{i=1}^n (y_i-\hat y_i)^2\;,
\end{equation}
where $y_i$ denotes the target and $\hat y_i$ the model prediction. 
It can be derived from a MLE loss with Gaussian likelihood with a universal width $\sigma$,
\begin{equation}
p(\mathcal{D}|\theta) 
= \prod_{i=1}^n G(y_i | \hat{y}_i, \sigma)
= \prod_{i=1}^n \frac{1}{\sqrt{2\pi}\sigma} 
  \exp\left(-\frac{(y_i - \hat{y}_i)^2}{2\sigma^2}\right)\;,
\end{equation}
so that the negative log-likelihood reduces to
\begin{equation}
-\log p(\mathcal{D}|\theta) 
= \frac{1}{2\sigma^2}\sum_{i=1}^n (y_i - \hat{y}_i)^2 + \text{const}\;,
\end{equation}
which, averaged over the dataset and ignoring the constant prefactor $1/\sigma^2$, gives $L_{\mathrm{MSE}}$.

\item[Binary cross-entropy:]
For binary classification with labels $y_i\in\{0,1\}$ and predicted probabilities $\hat p_i$, a common loss is the binary cross-entropy. 
It can be derived from a Bernoulli likelihood,
\begin{equation}
p(\mathcal{D}|\theta) 
= \prod_{i=1}^n \hat p_i^{\,y_i}(1-\hat p_i)^{1-y_i}\;,
\end{equation}
so that the negative log-likelihood gives
\begin{equation}
-\log p(\mathcal{D}|\theta) 
= -\sum_{i=1}^n
\left[
y_i \log \hat p_i + (1-y_i)\log(1-\hat p_i)
\right]\;,
\end{equation}
which, averaged over the dataset, defines the binary cross-entropy loss,
\begin{equation}
L_{\mathrm{BCE}}
=
-\frac{1}{n}\sum_{i=1}^n
\left[
y_i \log \hat p_i + (1-y_i)\log(1-\hat p_i)
\right]\;.
\end{equation}

\item[Categorical cross-entropy:]
For multi-class classification with one-hot encoded targets $y_{ik}$, \ie $y_{ik}=1$ if example $i$ belongs to class $k$ and $y_{ik}=0$ 
otherwise, and predicted class probabilities $\hat p_{ik}$, the natural generalization is
\begin{equation}
L_{\mathrm{CCE}}
=
-\frac{1}{n}\sum_{i=1}^n \sum_{k=1}^K
y_{ik}\log \hat p_{ik}\;.
\end{equation}

\item[Evaluation metrics:]
The choice of evaluation metric depends on the task. For regression, 
one often reports the mean squared error, mean absolute error, or 
coefficient of determination. For classification, predictions are 
first categorised as true positives (TP, correctly identified signal), 
true negatives (TN, correctly identified background), false positives 
(FP, background misidentified as signal), and false negatives 
(FN, signal misidentified as background), from which common metrics 
such as accuracy, precision, recall, ROC curves, and the area under 
the ROC curve (AUC) are derived.

\item[Accuracy:]
For binary classification, the accuracy is the fraction of correctly classified examples,
\begin{equation}
\text{Accuracy}
=
\frac{\text{TP}+\text{TN}}
     {\text{TP}+\text{TN}+\text{FP}+\text{FN}}\;.
\end{equation}

\item[Precision and recall:]
Precision and recall are defined as
\begin{equation}
\text{Precision}
=
\frac{\text{TP}}{\text{TP}+\text{FP}},
\qquad
\text{Recall}
=
\frac{\text{TP}}{\text{TP}+\text{FN}}\;.
\end{equation}
Recall is also referred to as the true positive rate or sensitivity.

\item[Sensitivity and specificity:]
The sensitivity and specificity are given by
\begin{align}
\text{Sensitivity}
&=
\frac{\text{TP}}{\text{TP}+\text{FN}}
\equiv \text{TPR}, \notag\\
\text{Specificity}
&=
\frac{\text{TN}}{\text{TN}+\text{FP}}
\equiv 1-\text{FPR}\;.
\end{align}

\item[ROC curve and AUC:]
The receiver operating characteristic (ROC) curve plots the true positive rate against the false positive rate as the classification threshold is varied. The area under this curve (AUC) provides a threshold-independent summary of classifier performance.

\item[Efficiency and background rejection:]
In many physics applications one quotes the signal efficiency
and background rejection,
\begin{equation}
\text{Efficiency}
=
\frac{N_{\text{selected}}^{\text{signal}}}
     {N_{\text{total}}^{\text{signal}}},
\qqquad
\text{Rejection}
=
1-
\frac{N_{\text{selected}}^{\text{bkg}}}
     {N_{\text{total}}^{\text{bkg}}}\;.
\end{equation}

\item[Purity:]
In classification and event-selection contexts, the purity of a selected sample is the fraction of selected events that truly belong to the target class. For a signal selection, this is given by
\begin{equation}
\text{Purity}
=
\frac{\text{TP}}{\text{TP}+\text{FP}}\;,
\end{equation}
which in binary classification coincides with the precision. Note that purity depends not only on the classifier performance, but also on the relative abundance of signal and background in the sample; a highly performant classifier may still yield low purity if the signal is rare.

\end{description}
Supervised learning shares many conceptual features with classical statistical analysis, including estimation, discrimination, and uncertainty assessment, while relying on flexible data-driven function classes that can adapt to 
complex and high-dimensional structure.

\subsection{Weak and Semi-supervised Learning}
\label{sec:semisupervised}

In many physics applications, fully labelled datasets are expensive, incomplete, or fundamentally unavailable. This is common, for example, when labels rely on imperfect simulation, when only aggregate information is known, or when the phenomenon of interest is too rare for large labelled samples to exist. Weak and semi-supervised learning address such settings.

\begin{description}

\item[Weak supervision:]
In weakly supervised learning, the available labels are noisy, biased, indirect, or only partially informative about the true target. This may occur, for example, when simulated labels are transferred to real data, when labels are assigned only at the level of groups of events, or when proxy observables are used instead of the quantity of true interest.

\item[Typical weak-supervision strategies:]
Common methods include label correction or smoothing, loss reweighting according to label confidence, and multiple-instance learning, in which labels are known only for sets or bags of examples rather than for each instance individually.

\item[Semi-supervised learning:]
Semi-supervised learning combines a small labelled dataset with a larger unlabelled one. The basic idea is to use the labelled data to anchor the task while exploiting the unlabeled data to learn a more robust representation or decision boundary.

\item[Typical semi-supervised strategies:]
Common approaches include self-training with pseudo-labels, confidence-thresholded pseudo-labeling, consistency regularization under perturbations of the input, and probabilistic generative models such as semi-supervised variational autoencoders.

\end{description}
In astrophysics, cosmology, and particle physics, such methods are increasingly used in settings where only a small subset of data can be annotated reliably, or where the target concept itself is partly theory-dependent. They are therefore likely to play an increasingly important role in future data analysis pipelines.

\subsection{Unsupervised Learning}
\label{sec:unsupervised}

Unsupervised learning refers to tasks in which no explicit target labels are provided. Instead, the aim is to identify structure in the input data, compress it into useful representations, or characterize atypical patterns. This is especially relevant in physics when labels are unavailable, ambiguous, or intrinsically ill-defined.

\begin{description}

\item[Clustering:]
Clustering methods group similar samples without supervision, thereby revealing latent classes, phases, or substructures in the data. Common examples include $k$-means clustering, hierarchical clustering, spectral clustering, and Gaussian mixture models.

\item[Dimensionality reduction:]
Dimensionality-reduction methods seek a lower-dimensional representation of high-dimensional data while preserving as much relevant structure as possible. Classical examples include principal component analysis, while non-linear approaches include manifold learning and autoencoders.

\item[Density estimation and generative models:]
A central unsupervised task is to learn the probability distribution $p(x)$ underlying the data, either explicitly through density estimation or implicitly through a generative model. 
Common approaches include normalizing flows, variational autoencoders, and diffusion models. These models are discussed in more detail in \cref{sec:generative} and in \emph{VERaiPHY article~5}~\cite{Diefenbacher:2026kki}.

\item[Anomaly detection:]
Anomaly-detection methods aim to identify rare or unusual samples that deviate significantly from the bulk of the data distribution. In physics, this is relevant for the search for unexpected phenomena, rare event classes, or detector pathologies. Common approaches include density estimation, one-class classification, clustering-based outlier scores, and reconstruction-based methods using, for example, autoencoders or normalizing flows. A detailed discussion is provided in \emph{VERaiPHY article~6}~\cite{Amram:2026vkc}.

\end{description}

\clearpage
\section{Common Machine Learning Model Families}

A wide variety of ML model families have been developed for tasks such as classification, regression, clustering, dimensionality reduction, and generative modelling. These models differ in their mathematical formulation, expressiveness, interpretability, computational cost, and ability to represent uncertainty.

In this section, we briefly review the most common classes of models used in scientific ML. While ML architectures play a central role in many current applications, classical and probabilistic approaches are still used in settings where interpretability, or principled uncertainty quantification are limiting current neural network implementations. The appropriate model choice depends not only on predictive performance, but also on the structure of the data, the scientific objective, and the computational and experimental constraints of the problem at hand.

\subsection{Classical Machine Learning Models}
\label{sec:classical}

Classical machine learning methods often provide strong baselines and are used in scientific applications in cases where valid constraints do not allow deep neural networks to unfold their power.

\begin{description}

\item[Linear and logistic regression:]
Linear regression models a continuous target as a linear function of the input features, while logistic regression models the probability of a binary outcome by passing a linear predictor through the logistic (sigmoid) function. Despite their simplicity, such models often provide robust and interpretable baselines.

\item[Support vector machines:] 
Support vector machines (SVMs) construct a decision boundary by maximizing the distance to the nearest training examples on either side, known as support vectors. Through kernel methods, they can also represent non-linear decision surfaces in implicitly mapped feature spaces.

\item[Decision trees and ensembles:]
Decision trees recursively partition the feature space into regions in which the target variable takes approximately constant values, assigning a single prediction to each region. Ensemble methods such as random forests and boosted decision trees improve predictive performance and robustness by combining many trees. In physics, boosted decision trees remain particularly popular for tabular classification problems.

\item[$k$-nearest neighbors:]
The $k$-nearest-neighbors ($k$-NN) method predicts targets from the labels or values of nearby training examples in feature space. It is conceptually simple and non-parametric, but becomes less effective in high dimensions.

\item[Classical unsupervised methods:]
Common unsupervised techniques include $k$-means clustering, density-based clustering such as DBSCAN, principal component analysis (PCA), and nonlinear embedding methods such as t-SNE. These methods are frequently used for exploratory data analysis, visualization, and compression.

\item[Gaussian processes:]
Gaussian processes (GPs) define a distribution over functions,
\begin{equation}
f(x)\sim \mathcal{GP}(m(x),k(x,x'))\;,
\end{equation}
where $m(x)$ is a mean function and $k(x,x')$ a covariance kernel. For training data $\mathcal{D}=\{(x_i,y_i)\}_{i=1}^n$, the predictive distribution at a test point $x_*$ is Gaussian, with mean and variance
\begin{align}
\mu_*
&=
k_*^\top (K+\sigma_n^2 I)^{-1}y\;, \notag\\
\sigma_*^2
&=
k(x_*,x_*) - k_*^\top (K+\sigma_n^2 I)^{-1}k_*\;,
\end{align}
where $K$ is the kernel matrix over the training inputs and $k_*$ denotes the vector of kernel evaluations between $x_*$ and the training points. GPs are especially useful in low-data regimes and when calibrated uncertainty estimates are important.

\end{description}

\subsection{Neural Networks}
\label{sec:nn}

Neural networks are flexible, parameterized function approximators built from compositions of affine transformations and nonlinear activation functions. They form the backbone of modern ML and are especially powerful for high-dimensional data.

\begin{description}

\item[Computational units:]
A basic neural-network unit computes a weighted sum of its inputs followed by a nonlinear activation,
\begin{equation}
y = \phi\left(\sum_{i=1}^d w_i x_i + b\right)\;,
\end{equation}
where $w_i$ are weights, $b$ is a bias, and $\phi$ is an activation function.

\item[Activation functions:]
Activation functions introduce the nonlinearity that gives neural networks their expressive power. Without them, a stack of linear layers would be equivalent to a single linear transformation, so increasing depth alone would not enable the model to learn genuinely non-linear relationships. Common activation functions include the sigmoid, hyperbolic tangent, rectified linear unit (ReLU), leaky ReLU, softplus, and softmax. Their choice influences optimization behaviour, expressiveness, and output interpretation.

\item[Multilayer perceptrons:]
Multilayer perceptrons (MLPs), also called feedforward neural networks, consist of stacked fully connected layers. They are general-purpose approximators but do not explicitly exploit spatial, sequential, or relational structure in the input.

\item[Convolutional neural networks:]
Convolutional neural networks (CNNs) are designed for data with local spatial structure, such as images. By using shared convolutional filters and pooling operations, they efficiently extract translation-sensitive or approximately translation-invariant features.

\item[Recurrent neural networks:]
Recurrent neural networks (RNNs) are designed for sequential data and maintain a hidden state that evolves along the sequence.
Variants such as long short-term memory (LSTM) networks and gated recurrent 
units (GRUs) improve the modelling of long-range dependencies; more recent 
extensions such as xLSTM further improve scalability and expressiveness.
In cosmology and astrophysics, such architectures are particularly relevant for the analysis of time-series data, such as light curves, gravitational wave signals, or transient events.

\item[Graph neural networks:]
Graph neural networks (GNNs) operate on graph-structured data by iteratively 
aggregating information from connected nodes or edges, where the graph topology 
may range from sparse local connectivity to fully connected graphs as in 
transformers. They are motivated by the prevalence of point-cloud and 
relational data in physics, and are particularly suitable for irregular 
structures such as particle clouds, detector hits, or sets of physics objects 
without a natural ordering or grid structure.

\item[Residual and dense architectures:]
Deep neural networks may suffer from optimization difficulties as depth increases. Residual networks alleviate this by learning corrections to the identity map,
\begin{equation}
y = \mathcal{F}(x,\theta) + x\;,
\end{equation}
while densely connected networks promote feature reuse by passing information from many earlier layers to later ones. Such architectures are widely used in deep-learning applications requiring large model depth.

\item[Transformers:]
Transformers are neural architectures based on self-attention mechanisms, which allow pairwise interactions between all elements of a sequence or set to be modelled directly. A central building block is the scaled dot-product attention mechanism,
\begin{equation}
\mathrm{Attention}(Q,K,V)
=
\mathrm{softmax}\!\left(\frac{QK^\top}{\sqrt{d_k}}\right)V\;.
\end{equation}
Here $Q$, $K$, and $V$ denote the query, key, and value matrices respectively, obtained by linear projections of the input, and $d_k$ is the dimension of the key vectors. Self-attention is the special case in which queries, keys, and values all derive from the same input sequence.

\item[Bayesian neural networks:]
Bayesian neural networks (BNNs) extend standard neural networks by treating their parameters $\theta$ as random variables with prior distributions, rather than as fixed but unknown quantities. Given a dataset $\mathcal{D}$, inference is based on the posterior
\begin{equation}
p(\theta|\mathcal{D})
=
\frac{p(\mathcal{D}|\theta)\,p(\theta)}{p(\mathcal{D})}\;.
\end{equation}
Predictions for a new input $x_*$ are then obtained by averaging over this posterior,
\begin{equation}
p(y_*|x_*,\mathcal{D})
=
\int \d\theta\, p(y_*|x_*,\theta)\,p(\theta|\mathcal{D})\;.
\end{equation}
This replaces a single point prediction by a predictive distribution and enables principled uncertainty quantification, including epistemic uncertainty associated with limited knowledge of the model parameters and, when included in the likelihood, aleatoric uncertainty associated with intrinsic data variability.
The exact posterior $p(\theta|\mathcal{D})$ is typically intractable for modern neural networks because of the high dimensionality and nonlinearity of the parameter space. Practical implementations therefore rely on approximate inference methods such as variational inference, Monte Carlo dropout, or MCMC; a detailed discussion is provided in \emph{VERaiPHY article~2}~\cite{Haussmann:2026gbi}.

\item[Physics-informed neural networks:]
Physics-informed neural networks (PINNs) incorporate known physical constraints, often in the form of differential equations, directly into the loss function. For a differential operator $\mathcal{N}$ and network output $u_\theta(x,t)$, one typically minimizes a loss of the form
\begin{equation}
L(\theta)
=
L_{\mathrm{data}}(\theta)
+
\lambda\,L_{\mathrm{phys}}(\theta)\;,
\end{equation}
where $L_{\mathrm{phys}}$ penalizes violations of the governing equations and is evaluated using automatic differentiation. PINNs are particularly useful when data are sparse but strong physical structure is known. 
By embedding differential equations into the loss, they can in principle extrapolate beyond the training data into unseen feature space regimes, a capability that is generally not available to purely 
data-driven models. 
However, such extrapolation should always be validated and accompanied by uncertainty estimates.

\end{description}
These architectures differ mainly in how they encode structure in the input data. Their success in physics often depends on how well the architecture matches the geometry, symmetries, or correlations of the underlying problem.

\subsection{Generative Models}
\label{sec:generative}

Generative models encode a probability distribution that captures the structure of a dataset, enabling the generation of new samples and, in some cases, explicit likelihood evaluation. They are increasingly important in scientific ML, for example in fast simulation, density estimation, data augmentation, and likelihood-free inference.

\begin{description}

\item[Normalizing flows:]
Normalizing flows are generative models with explicit likelihoods, constructed by transforming a simple base distribution into a more complex target distribution through a sequence of invertible mappings,
\begin{equation}
x = f_K\circ f_{K-1}\circ \cdots \circ f_1(z)
\qquad
z\sim p(z)\;,
\end{equation}
where $\circ$ denotes function composition.
The density of $x$ is obtained through the change-of-variables formula,
\begin{equation}
p(x)
=
p(z)\left|
\det\!\left(\frac{\partial z}{\partial x}\right)
\right|\;.
\end{equation}
Flows are especially useful when exact density evaluation and invertibility are required. A common approach to constructing the invertible mappings $f_k$ are coupling layers, which split the input $z = (z_a, z_b)$ into two parts and apply a conditioned transformation,
\begin{equation}
x_a = z_a\;, \qquad x_b = g(z_b;\, f_\theta(z_a))\;,
\end{equation}
where $f_\theta$ is an arbitrary neural network with trainable parameters $\theta$ and $g$ is an invertible transformation conditioned on $z_a$. 
Since $x_a=z_a$ is passed through unchanged, the Jacobian is block triangular and its determinant reduces to that of $g$ alone, making it computationally efficient. 
Affine coupling layers are a common special case, where
\begin{equation}
g(z_b;\, f_\theta(z_a)) = z_b \odot \exp(s_\theta(z_a)) + t_\theta(z_a)\;,  
\end{equation}
where $\odot$ denotes the element-wise (Hadamard) product.

\item[Autoregressive models:]
Autoregressive models factorize the joint distribution using the chain rule,
\begin{equation}
p(x)=\prod_{i=1}^d p(x_i|x_1,\dots,x_{i-1})\;.
\end{equation}
This permits exact likelihood computation, but generation is sequential and may therefore be computationally expensive.

\item[Variational autoencoders:]
Variational autoencoders (VAEs) are latent-variable models where the data distribution is represented as
\begin{equation}
p(x)
=
\int \d z\, p(x|z)\,p(z)\;.
\end{equation}
Since the exact latent posterior is generally intractable, VAEs optimize the evidence lower bound (ELBO),
\begin{equation}
\log p(x)
\ge
\mathbb{E}_{q_\phi(z|x)}[\log p_\theta(x|z)]
-
D_{\mathrm{KL}}\!\big(q_\phi(z|x)\|p(z)\big)\;.
\end{equation}
They provide a principled framework for representation learning and generative modelling with latent structure.

\item[Generative adversarial networks:]
Generative adversarial networks (GANs) are implicit generative models in which a generator $G(z)$ and discriminator $D(x)$ are trained in competition. A standard objective is
\begin{equation}
\min_G \max_D\;
\mathbb{E}_{x\sim p_{\mathrm{data}}}[\log D(x)]
+
\mathbb{E}_{z\sim p(z)}[\log(1-D(G(z)))]\;.
\end{equation}
GANs are capable of producing high-fidelity samples, but training can be unstable and may suffer from mode collapse.

\item[Diffusion models:]
Diffusion models generate data by learning to reverse a gradual stochastic noising process. 
In the forward process, Gaussian noise is added over a sequence of time steps; the model then learns the reverse denoising 
transitions. 
Common formulations include denoising diffusion probabilistic models (DDPMs), which operate in discrete time, and score-based generative models, which learn the score function $\nabla_x \log p(x)$ and sample via Langevin dynamics. 
The related framework of (conditional) flow matching (FM/CFM) learns a continuous vector field transporting a base distribution to the data distribution, and can be seen as a deterministic alternative with similar training objectives. 
These models have shown excellent sample quality and comparatively stable 
training, and are increasingly used in scientific generative modelling.

\end{description}
Generative models are of growing importance in physics, where they are used for fast detector simulation, surrogate modelling of expensive processes, likelihood-free inference, and the generation of structured synthetic datasets.

\subsection{Symbolic and Programmatic Machine Learning Models}

Some ML methods aim not only to predict accurately, but also to recover interpretable or even closed-form relationships from data. Such approaches are particularly appealing in the physical sciences, where understanding the learned relation may be as important as predictive performance. In an ML language, one might interpret a closed formula as an extremely regularized and compact latent representation, which often allows for successful extrapolation.

\begin{description}

\item[Symbolic regression:]
Symbolic regression searches over spaces of mathematical expressions in order to discover compact formulas that describe data. It is often used when one hopes to identify approximate laws, conserved quantities, or analytic relations between observables.

\item[Genetic programming and evolutionary methods:]
Genetic programming and related evolutionary algorithms explore model or expression spaces through mutation, recombination, and selection. They are widely used in symbolic modelling and in automated feature construction.

\item[Sparse identification of nonlinear dynamics:]
Sparse Identification of Nonlinear Dynamics (SINDy) seeks parsimonious differential equations from time-series data by expressing the dynamics in a library of candidate functions and selecting a sparse subset of active terms.

\item[Programmatic models:]
More broadly, some approaches aim to recover interpretable algorithmic or code-like representations from data. Such methods are relevant when the desired model is procedural, compositional, or constrained by a known symbolic structure.

\end{description}
These approaches are increasingly explored in scientific ML as tools for interpretable discovery and model compression.

\clearpage
\appendix
\section{Glossary of Symbols and Terms}
 
Several symbols are overloaded across physics, statistics, and ML.
Their intended meaning should therefore be inferred from context.
The glossary below is meant as a compact guide to the most common usages in this
article and throughout the VERaiPHY series.
Terms that are defined in the body of this article are followed by a section
reference; terms that are the primary focus of a dedicated VERaiPHY contribution
are marked with the label \emph{(VERaiPHY article N)} as a placeholder
to be updated once the corresponding articles are finalised.
 
\begin{table}[ht!]
\centering
\renewcommand{\arraystretch}{1.5}
\caption{Glossary of common physics terms and symbols.
Some symbols, such as $\mu$, $\sigma$, and $y$, are also used with different
meanings in statistics and ML.}
\begin{tabular}{L{3.0cm} L{11.5cm}}
\hline
\textbf{Symbol} & \textbf{PHYSICS term} \\
\hline
$\theta$, $\phi$ & Angular variables \\
$b$              & Background \\
$s$              & Signal \\
$\sigma$         & Cross section \\
$\varepsilon$    & Efficiency \\
$E$              & Energy \\
$L_{\mathrm{int}}$ & Integrated luminosity \\
$j$              & Jet \\
$\ell$           & Lepton (electron $e$, muon $\mu$, or tau lepton $\tau$) \\
$m$              & Mass (\eg $m_W$) \\
$p$              & Momentum \\
$\eta$           & Pseudorapidity \\
$y$              & Rapidity \\
$\mu$            & Signal strength \\
$\mathrm{CL}_\mathrm{s}$ & Modified-frequentist confidence-level statistic; see \cref{sec:hypo} \\
\hline
\end{tabular}
\end{table}

\clearpage
\renewcommand{\arraystretch}{1.5}
\begin{longtable}{L{3.0cm} L{11.5cm}}
\caption{Glossary of common statistical terms and symbols.} \label{tab:glossary_stats} \\
\hline
\textbf{Symbol} & \textbf{STATISTICS term} \\
\hline
\endfirsthead

\multicolumn{2}{r}{\small\textit{(continued)}} \\
\hline
\textbf{Symbol} & \textbf{STATISTICS term} \\
\hline
\endhead

\hline
\multicolumn{2}{r}{\small\textit{(continues on next page)}} \\
\endfoot

\hline
\endlastfoot
$p(x)$           & Probability distribution or density over $x$; see \cref{sec:basics} \\
$p(A | B)$       & Conditional probability; see \cref{sec:basics} \\
$H_0$            & Null hypothesis \\
$H_1$            & Alternative hypothesis \\
$p(\theta)$      & Bayesian prior for parameter $\theta$; see \cref{sec:basics} \\
$p(\theta | d)$  & Bayesian posterior for parameter $\theta$ given data $d$; see \cref{sec:basics} \\
$\rho(x,y)$      & Pearson correlation coefficient; see \cref{sec:descriptive} \\
$\mathcal{C}(\theta)$      & Coverage as a function of the true parameter value; see \cref{sec:intervals} \\
$d$              & Data (generic) \\
$\mathcal{L}$    & Likelihood; see \cref{sec:modelling} \\
$\mathcal{L}_{\text{prof}}$ & Profile likelihood; see \cref{sec:modelling} \\
$\lambda(\theta)$ & Profile likelihood ratio; see \cref{sec:modelling} \\
$\mathbb{E}_{x \sim p}[f(x)]$ & Expectation value of $f$ under distribution $p$ \\
$\bar{x}$        & Sample mean \\
$\hat{\theta}$   & Estimator or fitted parameter value \\
$\mathrm{Bias}(\hat{\theta})$ & Bias of an estimator; see \cref{sec:descriptive} \\
$\theta$         & Parameter of interest \\
$\nu$            & Nuisance parameter \\
$\boldsymbol{\Sigma}$         & Covariance matrix; see \cref{sec:descriptive} \\
$\sigma$         & Uncertainty or standard deviation \\
$\sigma_{\text{stat}}$ & Statistical uncertainty \\
$\sigma_{\text{syst}}$ & Systematic uncertainty \\
$n_{\mathrm{dof}}$ & Number of degrees of freedom; see \cref{sec:basics} \\
$z$              & Gaussian-equivalent significance; see \cref{sec:descriptive} \\
$T(\mathcal{D})$ & Test statistic; see \cref{sec:hypo} \\
$p$-value        & Tail probability under a specified hypothesis, most commonly the null hypothesis $H_0$; see \cref{sec:descriptive} \\
$\mathcal{N}(x|\mu,\sigma)$ & Normal distribution with mean $\mu$ and standard dev.~$\sigma$; see \cref{sec:basics} \\
$\mathcal{N}(\mu,\boldsymbol{\Sigma})$ & Multivariate normal dist.\ with mean $\mu$ and covariance $\boldsymbol{\Sigma}$; see \cref{sec:basics} \\
$\chi^2_N$       & $\chi^2$ distribution with $N$ degrees of freedom; see \cref{sec:basics} \\
$r_i = (d_i-t_i)/\sigma_i$ & Pull or normalized residual; see \cref{sec:gof} \\
$D_{\mathrm{KL}}(p\|q)$ & Kullback--Leibler divergence from $q$ to $p$; see \cref{sec:distances} \\
$D_{\mathrm{JS}}(p,q)$ & Jensen--Shannon divergence; see \cref{sec:distances} \\
$D_{\mathrm{TV}}(p,q)$ & Total variation distance; see \cref{sec:distances} \\
$W_1(p,q)$       & First Wasserstein distance; see \cref{sec:distances} \\
Kolmogorov--Smirnov test & Non-parametric test comparing an empirical cumulative distribution to a reference, or comparing two empirical distributions; see \cref{sec:gof} \\
MCMC & Markov Chain Monte Carlo; see \cref{sec:sampling} \\
Goodness of fit & Assessment of compatibility between observed data and a statistical model; see \cref{sec:gof}\\
Hypothesis testing & Procedure for assessing which of two competing hypotheses is better supported by data; see \cref{sec:hypo}\\

\end{longtable}

\renewcommand{\arraystretch}{1.5}
\begin{longtable}{L{3.0cm} L{11.5cm}}
\caption{Glossary of common machine-learning terms and symbols -- architectures and model families.} \label{tab:glossary_ml_arch} \\
\hline
\textbf{Symbol / Term} & \textbf{ML term — Architectures and Model Families} \\
\hline
\endfirsthead

\multicolumn{2}{r}{\small\textit{(continued)}} \\
\hline
\textbf{Symbol / Term} & \textbf{ML term — Architectures and Model Families} \\
\hline
\endhead

\hline
\multicolumn{2}{r}{\small\textit{(continues on next page)}} \\
\endfoot

\hline
\endlastfoot

Autoencoder (AE)         & Neural network trained to compress inputs into a low-dimensional latent space and reconstruct them; the bottleneck layer enforces dimensionality reduction; see \cref{sec:unsupervised} \\
AE bottleneck & The narrow intermediate layer of an autoencoder that acts as an information bottleneck, forcing the network to learn a compressed representation \\
Autoregressive model & Generative model that factorises the joint density sequentially; exact likelihood, sequential sampling; see \cref{sec:generative} \\
BNN & Neural network with prior distributions over weights; predictions are obtained by marginalising over the weight posterior; see \cref{sec:nn} \\
BERT               & Bidirectional Encoder Representations from Transformers; a large pre-trained language model based on masked self-attention; foundational to modern NLP \\
CNN                & Convolutional neural network; exploits local spatial structure via shared filters; see \cref{sec:nn} \\
Diffusion model    & Generative model that learns to reverse a gradual Gaussian noising process; high sample quality and stable training; see \cref{sec:generative} \\
CFM & Generative framework that trains a vector field to transport a simple base distribution to a target; \emph{(VERaiPHY article 5)}~\cite{Diefenbacher:2026kki} \\
Foundational model & Large-scale model (typically a transformer) pre-trained on broad data and fine-tuned for downstream tasks; examples include GPT-class and BERT-class models \\
GAN                & Generative adversarial network; generator and discriminator trained in competition; implicit generative model; see \cref{sec:generative} \\
GNN                & Graph neural network; operates on graph-structured data by aggregating over neighbours; see \cref{sec:nn} \\
Gaussian process & Distribution over functions defined by a mean and covariance kernel; exact posterior available in closed form for regression; see \cref{sec:classical} \\
Normalising flow & Neural network with tractable inverse; enables exact density evaluation via the change-of-variables formula; see \cref{sec:generative} \\
LLM & Autoregressive transformer trained on large text corpora; capable of in-context learning, instruction following, and multi-step reasoning \\
LSTM               & Long short-term memory network; recurrent architecture with gating mechanisms for long-range dependencies; see \cref{sec:nn} \\
MLP                & Multilayer perceptron; fully connected feedforward network; see \cref{sec:nn} \\
PINN & Neural network whose loss function incorporates residuals of governing differential equations; see \cref{sec:nn} \\
RNN                & Recurrent neural network; processes sequential data via a hidden state; see \cref{sec:nn} \\
Repulsive ensemble & Ensemble of neural networks trained with a repulsive interaction in parameter space to increase diversity and improve uncertainty coverage; \emph{(VERaiPHY article 2)}~\cite{Haussmann:2026gbi}\\
Surrogate model    & Cheap differentiable approximation to an expensive simulator or likelihood; used for fast inference and optimisation; \emph{(VERaiPHY article 4)}~\cite{Dax:2026qsm} \\
Transformer        & Architecture based on multi-head self-attention; models pairwise interactions across all elements of a set or sequence; see \cref{sec:nn} \\
VAE                & Variational autoencoder; latent-variable generative model trained via the ELBO; see \cref{sec:generative} \\
VQ-VAE & Autoencoder with a discrete latent space defined by a learned codebook; used for tokenisation and compression \\

\end{longtable}

\renewcommand{\arraystretch}{1.5}
\begin{longtable}{L{3.0cm} L{11.5cm}}
\caption{Glossary of common machine-learning terms and symbols -- training concepts and optimisation.} \label{tab:glossary_ml_train} \\
\hline
\textbf{Symbol / Term} & \textbf{ML term — Training Concepts and Optimisation} \\
\hline
\endfirsthead

\multicolumn{2}{r}{\small\textit{(continued)}} \\
\hline
\textbf{Symbol / Term} & \textbf{ML term — Training Concepts and Optimisation} \\
\hline
\endhead

\hline
\multicolumn{2}{r}{\small\textit{(continues on next page)}} \\
\endfoot

\hline
\endlastfoot

$\mathcal{D}$    & Dataset; typically $\mathcal{D}=\{(x_i,y_i)\}_{i=1}^n$ in supervised learning \\
$\mathcal{X}$    & Input space \\
$\mathcal{Y}$    & Output or target space \\
$x$              & Input feature vector or observation \\
$y$              & Target variable: class label in classification or continuous value in regression \\
$\hat y$         & Model prediction for input $x$ \\
$z$              & Latent variable or compressed representation \\
$f_\theta(x)$    & Parametrised predictive model with parameters $\theta$ \\
$\theta$         & Model parameters (weights) \\
$L(y,\hat y)$    & Loss function comparing prediction and target; see \cref{sec:supervised} \\
$\nabla_\theta L$ & Gradient of the loss with respect to model parameters \\
$\eta$           & Learning rate \\
$\mathcal{B}$    & Mini-batch used in stochastic optimisation \\
$\mathcal{R}_{\rm emp}$ & Empirical risk; average loss over the training set \\
$\mathrm{CV}_k$ & $k$-fold cross-validation estimate; see \cref{sec:sampling} \\
Amortisation     & Strategy in which an inference network is trained once to map observations directly to posterior parameters, amortising the cost of repeated inference \\
Attention        & Mechanism that computes a weighted combination of values based on query-key similarity; central to transformers; see \cref{sec:nn} \\
Contrastive learning & Self-supervised objective that attracts representations of similar samples and repels those of dissimilar ones; used for learning without labels; \emph{(VERaiPHY article 9)}~\cite{VERaiPHY_rep} \\
Data augmentation & Training-time transformations of inputs used to increase effective dataset size and enforce invariances \\
Differentiable programming & Design principle in which the full pipeline (simulation, model, loss) is made differentiable, enabling end-to-end gradient-based optimisation \\
Dropout          & Regularisation technique that randomly zeroes units during training to reduce co-adaptation and improve generalisation \\
Early stopping   & Halting training when validation loss stops improving, to prevent overfitting \\
Inductive bias   & Assumptions built into a model architecture or training procedure that favour certain solutions; \eg translational invariance in CNNs \\
Loss function    & Scalar objective minimised during training; common examples are MSE, cross-entropy, and ELBO; see \cref{sec:supervised} \\
Regularisation   & Any technique that reduces overfitting by constraining model complexity; see \cref{sec:supervised} \\
Token            & Discrete unit of input to a language or sequence model; may represent a word, sub-word, patch, or physics object \\
Weight decay ($L^2$) & Regularisation penalty proportional to $\|\theta\|^2$; equivalent to a Gaussian prior on weights \\

\end{longtable}

\renewcommand{\arraystretch}{1.5}
\begin{longtable}{L{3.0cm} L{11.5cm}}
\caption{Glossary of common machine-learning terms and symbols -- methods, evaluation, and physics applications.} \label{tab:glossary_ml_methods} \\
\hline
\textbf{Symbol / Term} & \textbf{ML term — Methods, Evaluation, and Physics Applications} \\
\hline
\endfirsthead

\multicolumn{2}{r}{\small\textit{(continued)}} \\
\hline
\textbf{Symbol / Term} & \textbf{ML term — Methods, Evaluation, and Physics Applications} \\
\hline
\endhead

\hline
\multicolumn{2}{r}{\small\textit{(continues on next page)}} \\
\endfoot

\hline
\endlastfoot

AUC              & Area under the ROC curve; threshold-independent classifier performance summary; see \cref{sec:supervised} \\
ROC curve        & Receiver Operating Characteristic curve; plots true positive rate vs.\ false positive rate as threshold varies; see \cref{sec:supervised} \\
Active learning  & Iterative strategy in which the model queries an oracle for labels on the most informative unlabelled examples; reduces annotation cost; \emph{(VERaiPHY article 1)}~\cite{Lavie:2026psw}\\
Amplification    & Technique for enriching a dataset with rare signal events, \eg by reweighting or targeted generation, to improve sensitivity \\
Anomaly detection & Identification of samples that deviate significantly from the learned data distribution; see \cref{sec:unsupervised}; \emph{(VERaiPHY article 6)}~\cite{Amram:2026vkc} \\
Clustering       & Grouping of samples based on similarity without labels; see \cref{sec:unsupervised} \\
Conformal prediction & Distribution-free framework for constructing prediction sets with guaranteed marginal coverage at finite sample size; \emph{(VERaiPHY article 2)}~\cite{Haussmann:2026gbi}\\
Dimensionality reduction & Projection of high-dimensional data into a lower-dimensional representation; see \cref{sec:unsupervised} \\
Double descent   & Phenomenon in which test error decreases, then increases, then decreases again as model capacity grows, challenging classical bias-variance intuition \\
Ensemble         & Collection of independently trained models whose predictions are aggregated to improve accuracy and uncertainty estimation; \emph{(VERaiPHY article 2)}~\cite{Haussmann:2026gbi}\\
Few-shot learning & Learning from very few labelled examples, often by leveraging a pre-trained representation or meta-learned initialisation \\
FPD / KPD & Metrics for evaluating the quality of generated physics events by comparing feature distributions in a learned embedding space; analogues of the Fr\'{e}chet Inception Distance; \emph{(VERaiPHY article 5)}~\cite{Diefenbacher:2026kki} \\
Integrated gradient & Attribution method that assigns feature importance by integrating gradients along a path from a baseline to the input \\
Latent variable  & Unobserved variable in a probabilistic model whose posterior is inferred from data; central to VAEs and flows \\
Meta-learning    & Learning-to-learn; training a model on a distribution of tasks so that it can adapt rapidly to new tasks with few examples \\
Multi-modal learning & Learning from data of heterogeneous types (\eg text, images, tabular features) within a unified model \\
Mutual information & $I(X;Y) = D_{\mathrm{KL}}(p(x,y)\|p(x)p(y))$; measures statistical dependence between two variables; used in representation learning objectives \\
Natural language processing (NLP) & Field concerned with modelling and processing human language; increasingly relevant for scientific literature and data interfaces \\
Neural importance sampling & Use of a trained generative model as a proposal distribution within Monte Carlo integration to reduce variance; \emph{(VERaiPHY article 5)}~\cite{Diefenbacher:2026kki} \\
Optimal transport & Mathematical framework for comparing distributions via the minimal cost of transporting mass from one to another; includes Wasserstein distances; see \cref{sec:distances} \\
Overfitting      & Degraded generalisation caused by a model adapting too closely to training-set fluctuations; see \cref{sec:coreml} \\
Physics-informed AI & Design paradigm that embeds physical symmetries, conservation laws, or governing equations into model architecture or training \\
Prediction head  & Task-specific output layer appended to a pre-trained backbone; fine-tuned for a downstream classification or regression task \\
Representation learning & Learning of informative latent features from data, often without labels; see \cref{sec:coreml}; \emph{(VERaiPHY article 9)}~\cite{VERaiPHY_rep} \\
SHAP             & SHapley Additive exPlanations; attribution method based on game-theoretic Shapley values that assigns feature importance consistently; \emph{(VERaiPHY article 8)}~\cite{Gambhir:2026cly} \\
SBI & Simulation-based inference; likelihood-free inference using neural estimators of the posterior, likelihood, or likelihood ratio, trained on simulator outputs; \emph{(VERaiPHY article 4)}~\cite{Dax:2026qsm} \\
SDE & Stochastic differential equation driven by a noise process; governs forward and reverse dynamics in diffusion-based generative models \\
Surrogate model  & See \cref{tab:glossary_ml_arch} \\
Symmetries in AI & Incorporation of physical symmetry groups (\eg rotation, permutation, gauge invariance) into architecture design; improves data efficiency and equivariance; \emph{(VERaiPHY article 7)}~\cite{VERaiPHY_sym} \\
$t$-SNE          & $t$-distributed stochastic neighbour embedding; nonlinear dimensionality-reduction method for visualisation of high-dimensional data \\
UMAP             & Uniform Manifold Approximation and Projection; topology-based nonlinear dimensionality-reduction method; often faster than $t$-SNE with better global structure preservation \\
Validation       & Assessment of model performance on held-out data not used during training; see \cref{sec:coreml}; \emph{(VERaiPHY article 2)}~\cite{Haussmann:2026gbi}\\
Variational inference & Approximation of an intractable posterior by optimising over a family of simpler distributions, typically via the ELBO; see \cref{sec:generative} \\
Weak supervision & Training with noisy, indirect, or aggregate labels rather than exact instance-level supervision; see \cref{sec:semisupervised} \\
Mode collapse & Failure mode of generative models, often GANs, in which the generator produces samples covering only a subset of the true data distribution; see \cref{sec:generative} \\

\end{longtable}

\clearpage
\bibliography{refs}

\end{document}

%% file: incl_settings.tex
\usepackage[utf8]{inputenc} 
\usepackage[T1]{fontenc} 	
\usepackage[english]{babel} 

\usepackage[bitstream-charter]{mathdesign}
\usepackage{geometry} 		
\usepackage{microtype}
\usepackage{array}
\usepackage{amsmath} 		
\usepackage{mathtools} 		
\usepackage{float} 			
\usepackage{graphicx} 		
\usepackage{tabularx} 		
\usepackage{booktabs} 		
\usepackage{color, xcolor} 	
\usepackage{pdfpages} 		
\usepackage{extarrows} 		
\usepackage{multirow} 		
\usepackage{multicol} 		
\usepackage{enumitem} 		
\usepackage{xspace} 		
\usepackage{stackrel} 		
\usepackage{tikz} 			
\usepackage{braket} 		
\usepackage{bm} 			
\usepackage{tensor} 		
\usepackage{slashed} 		
\usepackage{siunitx} 		
\usepackage{lastpage} 		
\usepackage{cite} 			
\usepackage[normalem]{ulem} 
\usepackage{fontawesome} 	
\usepackage{tocloft} 		
\usepackage{titlesec} 		
\usepackage{doi} 			
\usepackage{hyperref} 		
\usepackage[most]{tcolorbox} 					
\usepackage[nameinlink, capitalize]{cleveref} 	
\usepackage[nottoc, notlot, notlof]{tocbibind} 	
\usepackage[ruled, vlined]{algorithm2e} 		
\usepackage{makecell}
\usepackage[makeroom]{cancel}
\usepackage{feynmf}

\makeatletter
\def\BState{\State\hskip-\ALG@thistlm}
\makeatother

\makeatletter
\@ifundefined{pdfoutput}{}{\DeclareGraphicsRule{*}{mps}{*}{}}
\makeatother

\makeatletter
\DeclareRobustCommand*{\bfseries}{%
   \not@math@alphabet\bfseries\mathbf
   \fontseries\bfdefault\selectfont
   \boldmath
}
\makeatother

\hypersetup{
	pdftitle={VERaIPHY},
	pdfauthor={Grosso et al.},
	colorlinks=true, 			
	linkcolor={red!50!black}, 	
	citecolor={blue!50!black}, 	
	urlcolor={blue!80!black} 	
} 

\DeclareSymbolFont{usualmathcal}{OMS}{cmsy}{m}{n}
\DeclareSymbolFontAlphabet{\mathcal}{usualmathcal}

\SetArgSty{textnormal}
\SetKwComment{Comment}{{\small\#}~}{}
\SetCommentSty{mycommfont}

\setitemize{itemsep=0pt, parsep=0pt} 				
\setenumerate{itemsep=0pt, parsep=0pt} 				
\setitemize{itemsep=2pt,topsep=2pt,parsep=0pt,partopsep=0pt,leftmargin=*}
\setenumerate{itemsep=0pt,topsep=2pt,parsep=0pt,partopsep=0pt,labelindent=3pt,leftmargin=*}
\newcolumntype{L}[1]{>{\raggedright\arraybackslash}p{#1}}
\setlist[description]{style=nextline}

\newlist{todolist}{itemize}{2}
\setlist[todolist]{label=$\square$}
\usepackage{pifont}

\usepackage{amsmath}
 
\usepackage{amsthm} 		
\theoremstyle{definition}

%% file: incl_shortcuts.tex
\definecolor{red_cb}{HTML}{e41a1c}
\definecolor{blue_cb}{HTML}{377eb8}
\definecolor{green_cb}{HTML}{4daf4a}
\definecolor{purple_cb}{HTML}{984ea3}
\definecolor{orange_cb}{HTML}{ff7f00}

\definecolor{EmeraldGreen}{HTML}{1ea78d}
\definecolor{EnglishRed}{HTML}{b02427}
\hypersetup{colorlinks=true,urlcolor=EmeraldGreen,citecolor=EmeraldGreen,linkcolor=EnglishRed}

\newcommand{\eg}{\text{e.g.}\xspace}
\newcommand{\ie}{\text{i.e.}\xspace}

\newcommand{\mwith}{\text{with}}
\newcommand{\mand}{\text{and}}

\newcommand{\qqquad}{\qquad\quad}

\def\d{\mathrm{d}}

\newcommand\one{\leavevmode\hbox{\small1\normalsize\kern-.33em1}}

\newcommand{\arXiv}[2][]{%
	\ifthenelse{\equal{#1}{}}%
	{\href{http://arxiv.org/abs/#2}{arXiv:#2}}%
	{\href{http://arxiv.org/abs/#2}{arXiv:#2~[#1]}}}

\def\slashchar#1{\setbox0=\hbox{$#1$}           
   \dimen0=\wd0                                 
   \setbox1=\hbox{/} \dimen1=\wd1               
   \ifdim\dimen0>\dimen1                        
      \rlap{\hbox to \dimen0{\hfil/\hfil}}      
      #1                                        
   \else                                        
      \rlap{\hbox to \dimen1{\hfil$#1$\hfil}}   
      /                                         
   \fi}

\newcommand{\tikznode}[2]{%
\ifmmode%
\tikz[remember picture,baseline=(#1.base),inner sep=0pt] \node (#1) {$#2$};%
\else
\tikz[remember picture,baseline=(#1.base),inner sep=0pt] \node (#1) {#2};%
\fi}

\def\mathswitchr#1{\relax\ifmmode{\mathrm{#1}}\else$\mathrm{#1}$\xspace\fi}
\def\mathswitch#1{\relax\ifmmode#1\else$#1$\xspace\fi}